\pdfoutput=1
\documentclass[12pt,preprint]{aastex}
\usepackage{graphicx,apjfonts,emulateapj5,onecolfloat5,dblfloatfix}
\usepackage{amssymb,mathrsfs,color,gensymb}
\usepackage{lineno}
\usepackage{longtable}
\usepackage{multicol}
\usepackage{blindtext}
\usepackage{ulem}
\usepackage{amsmath} 
\newfont{\rsfsten}{rsfs10 scaled 1200}
\newfont{\rsfsseven}{rsfs7 scaled 1200}
\newfont{\rsfsfive}{rsfs5 scaled 1200}

\slugcomment{Draft: \today, started May 12, 2021}

\shorttitle{}
\shortauthors{Zhao, Morris, Goss}
\begin{document}
\twocolumn[
\title{Detection of a  dense group of  hyper-compact radio sources in the central parsec of the Galaxy}
\author{
Jun-Hui Zhao\altaffilmark{1}, 
Mark R. Morris\altaffilmark{2} \& 
W. M. Goss\altaffilmark{3}
} 
\affil{$^1$Center for Astrophysics | Harvard-Smithsonian, 60
Garden Street, Cambridge, MA 02138, USA; jzhao@cfa.harvard.edu}
\affil{$^2$Department of Physics and Astronomy, University of California Los Angeles, Los Angeles, CA 90095}
\affil{$^3$NRAO, P.O. Box O, Socorro, NM 87801, USA}
\begin{abstract}
Using the JVLA, we explored the Galactic center (GC) with a resolution of 0.05" at 33.0 
and 44.6 GHz. We detected 64 hyper-compact radio sources (HCRs) in the central parsec. 
The dense group of HCRs can be divided into three spectral types: 38 steep-spectrum 
($\alpha\le-0.5$) sources; 10 flat-spectrum ($-0.5<\alpha\le0.2$) sources; and 17 
inverted-spectrum sources having $\alpha>0.2$, assuming $S\propto\nu^\alpha$. 
The steep-spectrum HCRs are likely represent a population of massive stellar 
remnants associated with nonthermal compact radio sources powered by neutron stars 
and stellar black holes. The surface-density distribution of the HCRs 
as function of radial distance ($R$) from Sgr~A* can be described as a steep 
power-law $\Sigma(R)\propto\/R^{-\Gamma}$, with $\Gamma=1.6\pm0.2$, along with 
presence of a localized order-of-magnitude enhancement in the range 0.1-0.3 pc. 
The steeper profile of the HCRs relative to that of the central cluster 
might result from the concentration massive stellar remnants by mass segregation 
at the GC. The GC magnetar SGR~J1745-2900 belongs to the inverted-spectrum sub-sample. 
We find that two spectral components present in the averaged radio spectrum 
of SGR~J1745-2900, separated at $\nu\sim30$ GHz. The centimeter-component is 
fitted to a power-law with $\alpha_{cm}=-1.5\pm0.6$. The enhanced millimeter-component 
shows a rising spectrum $\alpha_{mm}=1.1\pm0.2$. Based on the ALMA observations 
at 225 GHz, we find that the GC magnetar is highly variable on a day-to-day time scale, 
showing variations up to a factor of 6. Further JVLA and ALMA observations of 
the variability, spectrum, and polarization of the HCRs are critical for 
determining whether they are associated with stellar remnants.
\vskip 5pt
\noindent 
{{\it Unified Astronomy Thesaurus concepts:} {\color{black}  
Center of the Milky Way; [Galactic center (565)]; Interstellar medium (847);
Radio continuum emission (1340); Black holes (162); Pulsars (1306); 
Magnetars (992); Neutron stars (1108);  
Discrete radio sources (389); Radio transient sources (1358); Radio interferometry (1346) 
}}
\end{abstract}
]

\section{Introduction}

One of the outstanding open questions that has challenged astronomers for many years is the
``missing pulsar problem'': there are far fewer pulsars found toward the Galactic center (GC) 
than we could expect, given the formation rate of massive stars in the central molecular zone 
of the Galaxy implied by the relative abundance of massive stars produced at the GC over the past 
ten million years  \citep[e.g.,][]{don2012, lu2013, cla2021}.  This can in part be ascribed to the 
large foreground scatter-broadening at  longer radio wavelengths toward  the GC, which can 
lead to a large enough pulse broadening that the pulses become indistinguishable \citep{laz1998}. 
Several other reasons also complicate the discovery of GC pulsars, as detailed by \citet{eat2021}.  
However, the discovery of a magnetar associated with SGR J1745-2900, located just 3" from Sgr A*  
\citep{ken2013,mor2013,rea2013}, indicates that the effect of the scattering screen could be up 
to three orders of magnitude smaller than had previously been expected \citep{spi2014,bow2014}. 
Consequently, the question remains: why haven't more pulsars been seen toward the Galactic center? 
Because massive stars clearly form in abundance at the GC, and because dynamical friction should 
cause the more massive stellar remnants to be concentrated there, neutron stars should be abundant 
and continuously produced in the GC region ($R\sim0.5$ pc) 
\citep{bah1976,mor1993,mir2000,pfa2004,ale2009,mer2010,ant2012,ale2017}.

One obvious answer to this question of where the pulsars are is that the number of ``windows'' 
in the scattering screen is quite small, so that most pulsars are still too scatter-broadened 
for their pulses to be detected at the wavelengths searched. Another, perhaps more interesting 
answer is that massive stars that form out of the rather highly magnetized interstellar 
medium of the GC \citep{mor2014} tend to themselves be rather strongly magnetized, and 
therefore leave strongly magnetized neutron star remnants. That is, pulsars formed near the 
GC could frequently be magnetars, which have short lifetimes as recognizable pulsars 
($\sim10^3 - 10^5$ yrs) because of their rapid spin-down rates \citep{har1999, esp2011, kas2017}. 
Such short lifetimes would limit the number of pulsars that could be detected at any one moment 
to a small number, although they could remain detectable as compact radio sources.
Radio continuum surveys of point sources can help distinguish these possibilities. 
We recently published a 5.5-GHz survey of GC compact  radio sources (GCCRs) within a 
radius of $\sim$7.5 arcmin (17 pc) of Sgr A* \citep{zmg2020}, and concluded that, 
of the 110 new compact radio sources observed down to a 10-$\sigma$ sensitivity 
limit of 70 $\mu$Jy, most of them fall within the high flux density tail of normal 
pulsars at the GC (our effort to decrease the 5.5-GHz flux density limit with existing, 
additional data is in progress). Of course, there are several other possible 
assignations for these sources; 82 of them are variable or transient and 42 have 
possible X-ray counterparts. 

Limited by the VLA angular resolution and confusion from the HII continuum emission from 
Sgr A West, the 5.5-GHz survey focussed primarily on regions lying beyond a radius of R $\sim$ 1 pc 
from Sgr A*, that is, on regions outside the circumnuclear disk. To take the next step in 
addressing the pulsar puzzle, we have recently surveyed the central $\pm$0.5 parsecs ($\pm$13") 
around Sgr A* at higher frequencies, using existing JVLA Ka and Q-band data and X-band observations 
in the A-array. The high-resolution JVLA observations at 33 and 44.6 GHz were used to search 
for hyper-compact (< 0.1") radio sources (HCRs) in Sgr A West, and to study their radio 
properties and distribution near Sgr A*. The motivation for going to higher frequencies 
in the context of constraining the magnetar population comes from the discovery by \cite{tor2017} 
that the spectrum of the magnetar near Sgr A*, SGR J1745-2900, rises at higher frequencies 
to a millimeter/submillimeter plateau. Another magnetar, 1E 1547.0$-$5408, has also been 
seen to display a spectrum rising at millimeter wavelengths \citep{chu2021}. If such a 
rising spectrum happens to be a general characteristic of magnetars, then this feature 
can be used to identify magnetar candidates with higher frequency observations, even if 
radio pulses are not detectable.

This paper is organized as follows: Section 2 describes the JVLA observations, data 
reduction and imaging procedures used for identifying HCRs within the central parsec.  
We also describe there our procedure for data reduction and imaging using archival ALMA data 
for measurements of SGR J1745-2900. Section 3 presents a catalog of the HCRs found within 
Sgr A West.  Three selected cases of HCRs are described in Section 4, including detailed
results on the radio spectrum and variability of the GC magnetar SGR J1745-2900, based on  
data from this paper and from prior publications. Section 5 shows a statistical analysis of 
HCRs by dividing them into three spectral types. Possible origins of the HCRs as well as 
massive stellar remnants as a consequence of the mass segregation in the central parsec 
are also discussed in section 5. Finally, section 6 summarizes our conclusions.

\begin{figure*}[!ht]
\centering
\includegraphics[angle=0,width=135.mm]{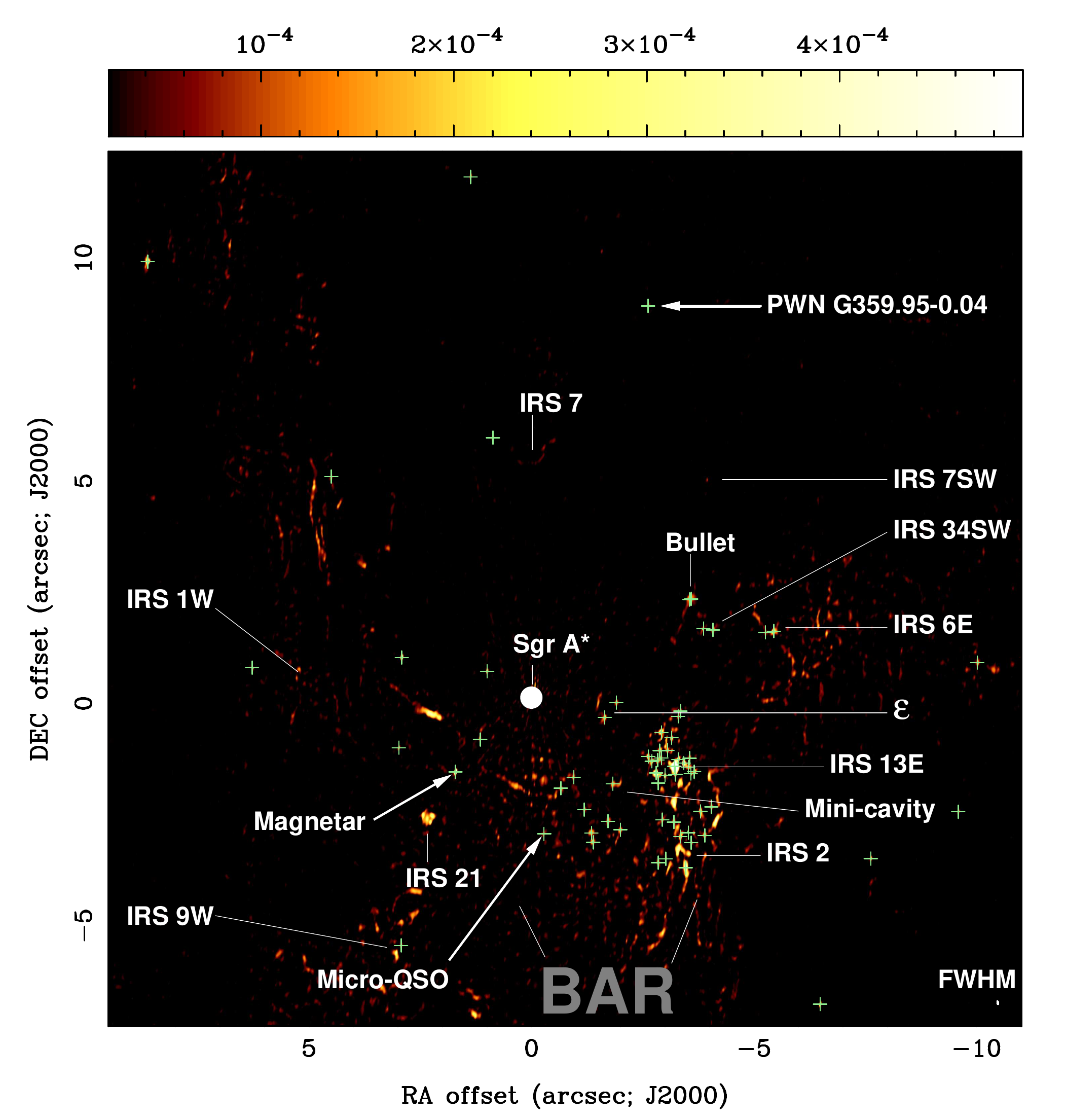}
\caption{Ka-band image of the central 0.8 parsec of the Galaxy. Sgr A* is indicated as 
a white filled circle, and all 64 HCRs that we have identified in this region are marked 
with green crosses. This 33-GHz image was made with baselines from 500 to 4,000 k$\lambda$, 
giving a FWHM beam of 0.08$\times$0.03" and an rms noise of 8 $\mu$Jy beam$^{-1}$. The 
color wedge scales intensity in units of Jy beam$^{-1}$. In addition to the marked HCRs, 
some extended features corresponding to fine-scale structures in the Northern Arm, Eastern Arm, 
and the ``mini-cavity'' of the Sgr A West HII region, which are present in this image, are 
also labelled. The position of Sgr A* is registered at RA(J2000)=17:45:40.0409, 
Dec(J2000)= $-$29:00:28.118 \citep{rei2014} in the equatorial coordinate system.
}
\label{fig1}
\end{figure*}

\begin{table*}[!ht]
\tablenum{1}
\footnotesize
\centering
\setlength{\tabcolsep}{1.7mm}
\caption{Log of JVLA datasets and images}
\begin{tabular}{lcccccccccc}
\hline\hline \\
&\multicolumn{6}{c}{\underline{~~~~~~~~~~~~~~~~~~~~~~~~~~~~~~~~~ UV data ~~~~~~~~~~~~~~~~~~~~~~~~~~~}}& 
&\multicolumn{3}{c}{\underline{~~~~~~~~~~~~~~~~~~~~~~~~ Images  ~~~~~~~~~~~~~~~~~~~~~~}} \\
{Project ID}&
{Array}&
{Band}&
{$\nu$}&{$\Delta\nu$~~~} &
{$\Delta t$}&
{HA range} &
{~~~~~~~~Epoch~~~~~~~~} & Weight & {\color{black}($\theta_{\rm maj}$,$\theta_{\rm min}$, 
${\rm PA}$)}& rms\\
{} &
{} &
{} &
{(GHz)} &
{(GHz)} &  &
{(sec)}&
{(day)} &
{(R)}   &
{(arcsec, arcsec, deg)}&
{($\mu$Jy beam$^{-1}$)}
\\
{(1)}&{(2)}&{(3)}&{(4)}&{(5)}&{(6)}&{(7)}&(8)&(9)&(10)&(11)\\
\hline \\
\vspace{5pt}
15A-293 &A &Q$^\dagger$& 44.6&8&3&$-0^h.6$ --- $+2^h.5$&2015-09-16&0&0.078, 0.032, 12&17\\
\vspace{5pt}
\dots &\dots&Ka$^\dagger$& 33.0&8&2&$-2^h.8$ --- $+0^h.3$&2015-09-11&$-$0.3&0.079, 0.031, $-11$&8\\
\vspace{5pt}
14A-346 &A &X$^\ddagger$& 9.0&2&2&$-3^h.4$ --- $+3^h.3$&2014-04-17&0&0.36, 0.15, $-6$&4\\
\vspace{5pt}
19B-289 &A &X$^\ddagger$& 9.0&2&2&$+0^h.6$ --- $+2^h.6$&2019-09-21&0&0.36, 0.15, $-6$&7\\
\vspace{5pt}
20B-203 &A &X$^\ddagger$& 9.0&2&2&$-0^h.3$ --- $+2^h.9$&2020-11-20&0&0.36, 0.15, $-6$&5\\
\vspace{5pt}
\dots   &\dots &\dots&\dots&\dots&\dots&$-1^h.5$ --- $+1^h.7$&2020-12-04&\dots&\dots&\dots\\
\hline
\end{tabular}\\
\vspace{2pt}
\begin{tabular}{p{0.85\textwidth}}
{
\tiny
(1) JVLA program code of PI: Mark Morris for 19B-289 and 20B-203; PI: Farhad Yusef-Zadeh for 14A-346 and 15A-293.
(2) Array configurations.
(3) JVLA band code; "X","Ka", and "Q" stand for the VLA bands in the ranges of 
$8.0-12.0$ GHz, $26.5-40.0$ GHz and $40.0-50.0$ GHz
(https://science.nrao.edu/facilities/vla/docs/manuals/oss2013B/performance/bands).
(4) observing frequencies at the observing band center.
(5) bandwidth.
(6) integration time.
(7) hour angle (HA) range for the data.
(8) date corresponding to the image epoch.
(9) robustness weight parameter.
(10) FWHM of the synthesized beam.
(11) rms noise of the image. 
$^\dagger$Correlator setup: 64 channels in each of 64 subbands with channel width of 2 MHz
$^\ddagger$Correlator setup: 64 channels in each of 16 subbands with channel width of 2 MHz. 
}
\end{tabular}
\end{table*}

\section{Observations,  data \& imaging}

Following the 5.5 GHz JVLA A-array survey of compact radio sources within a radius of  
7.5 arcmin (17 pc) in the Galactic center region, we have carried out a search for 
compact radio sources within a radius of 13" (0.5 pc) based on the existing JVLA 
high-resolution data at 44.6 GHz (Q-band) and 33.0 GHz (Ka-band) as well as our 
recent JVLA A-array observations at 9 GHz. The high-resolution and sensitive VLA
observations at high radio frequencies are crutial in detections of HCRs at a level 
of 100 $\mu$Jy to a few mJy in the vicinity of Sgr A*. 

\subsection{JVLA data, calibration \& imaging}

New JVLA observations in the A configuration were carried out on 2019-9-21, 
2020-11-20 and 2020-12-4 at 9 GHz, in addition to the observation on 2014-04-17 
that was used to image the Sgr A West filament \citep{mzg2017}. The X-band observations 
were all carried out with the VLA standard correlator setup for wideband continuum 
covering 2 GHz bandwidth, produced from the 8-bit sampler. We also acquired the 
higher-resolution NRAO archival data observed with the JVLA in the A-array at 33 and 
44.6 GHz on 2015-9-16 and 2015-9-11, respectively. The Q- and Ka-band observations 
were using the 3-bit sampler, producing 64 subbands and covering a total of 8 GHz 
bandwidth. All the observations were pointed at a sky position\footnote{RA(J2000) =
17:45:40.0383, Dec(J2000) = $-$29:00:28.069}, very near Sgr A*. Table 1 summarizes 
the six sets of uv data  (columns 1 - 8). The data reduction was carried out using 
the CASA\footnote{http://casa.nrao.edu} software package of the NRAO. The standard 
calibration procedure for JVLA continuum data was applied. 3C 286 (J1331+3030) was 
used for corrections for delay, bandpass and flux-density scale. 

J1733-1304 (NRAO 530) and J1744-3116 were used for complex gain calibrations. 
In addition, corrections for the time variation of the bandpass  across each baseband 
due to residual delays were determined using NRAO 530 based on the model discussed 
in \cite{zmg2019} . The accuracy of the flux-density scale at the JVLA is 3\%$-$5\%, 
limited by the uncertainty of the flux density of the primary calibrator, Cygnus A \citep{per17}.
Following the procedure for high dynamic range (DR) imaging that we developed recently
\citep{zmg2019} and applying it to the Sgr A data with CASA, we further corrected for the residual
errors in phase. The integration time of the calibrated Q and Ka-band data was averaged 
into 30-sec time bins, so that the intensity loss at the 13'' outer radius of the 
search region is less than 10\% due to the smearing effect caused by the Earth's spin. 
After correcting for the residual delay, we also binned the spectral channel data to channel
widths of 32 and 16 MHz for the Q- and Ka-band data, respectively, to ensure that the 
intensity loss for a point source caused by bandwidth smearing is less than 10\% at 
the radius of 13".  The short baseline data ($<500{\rm k}\lambda$) were filtered out in 
order to avoid contamination by the extended emission in Sgr A West on scales $>0.4$". 
The Ka-band image shown in Fig. 1 is made with CASA task $CLEAN$ from the baseline data between 
500-4,000 k$\lambda$, achieving an rms of 8 $\mu$Jy beam$^{-1}$ with a FWHM beam 
of 0.079"$\times$0.031 ($-11^{\rm o}$). This high-resolution image shows 
numerous compact radio sources in the central parsec region. 

For the X-band data, after corrections for the residual errors, the visibility data were 
averaged to a time bin of 16 sec while the original channel width of 2 MHz was kept. 
We made images at 9 GHz using the multi-frequency synthesis algorithm \citep[MFS:][]{rau11}  
with the 1024 spectral channels covering the 2 GHz bandwidth. Also we filtered out 
the short baselines ($<100$ k$\lambda$), and constructed two 9-GHz images for both the
2019 and 2020 epochs. The second epoch image was made with the two observations on 
2020-11-20 and 2020-12-04.  Hereafter, we use the mean epoch, 2020-11-27, for this image.  
The rms noise for the 2020-11-27 X-band  image is 5 $\mu$Jy beam$^{-1}$. We reconstructed 
the 2014-04-17 image with the calibrated X-band A-array data \citep{mzg2017} by using 
only the long-baseline data ($>100$ k$\lambda$). The rms noise for the resulting 
2014-04-17 X-band image is 4 $\mu$Jy beam$^{-1}$.

The specified parameters for the final images at Q-, Ka- and X-bands are 
summarized in Table 1 from columns 9 to 11.

\subsection{ALMA data, calibration \& imaging}
We acquired archival data from the Atacama Large Millimeter Array (ALMA), 
observed by \cite{tsu2019} at 225.75 GHz in Cycle 5 (2017.1.00503.S). 
Following the ALMA CASA guide for Cycle 5 data reduction, we executed  
the pipeline script ``{\bf scriptForPI.py}'' to produce calibrated 
ALMA data in CASA Measurement Set format. The ALMA datasets are composed 
of ten 1-hr observations of IRS 13E in the array configuration of C43-10  
within a two-week interval between 2017-10-6 and 2017-10-20. The observing 
field covers Sgr A* and the magnetar SGR J1745-2900 in addition to IRS 13E. 
The pipelined ALMA images appear to be marred by severe residual errors. Both 
IRS 13E and SGRA J1745-2900 were buried in the sidelobes and artifacts produced 
by the residual dirty beam. Then, following the recipe for dynamic range imaging 
with wideband data \citep{zmg2019}, we corrected the residual errors by 
utilizing the antenna-based closure relations \citep{tms2017} and constructed 
a refined ALMA image from the data observed at the first epoch on 2017-10-6 
as a trial case using CASA task $tCLEAN$ with the MFS algorithm and robust 
weighting with Briggs parameter R=0 \citep{bri1995}.  An rms noise of 
$\sigma=20~\mu$Jy beam$^{-1}$ was achieved, with a FWHM beam of 0.024"$\times$0.017" 
(82$^{\rm o}$). The magnetar SGR J1745-2900 was successfully detected with a S/N 
ratio of 70.  We then made 2D-Gaussian fits to both Sgr A* and the magnetar, 
finding flux densities of 3.031$\pm$0.013 Jy and 1.41$\pm$0.08 mJy at 225 GHz 
for Sgr A* and the magnetar, respectively. Fig. 2a shows the ALMA continuum image 
of the magnetar at 225.75 GHz.  Based on the procedure and input parameters for 
the CASA tasks that were used for the trial case, we coded the detailed CASA reduction 
steps for the ALMA data into a CASA-Python script. Using this script for corrections 
of residual errors, we further processed the datasets from all ten epochs observed 
on 2017-10-6, 2017-10-7, 2017-10-9, 2017-10-10, 2017-10-11, 2017-10-12, 2017-10-14, 
2017-10-17, 2017-10-18, and 2017-10-20. The flux densities for the magnetar were 
determined by fitting a 2D-Gaussian and are tabulated in Table 3 (see Section 4).

\begin{figure}[!h]
\centering
\includegraphics[angle=0,width=82.5mm]{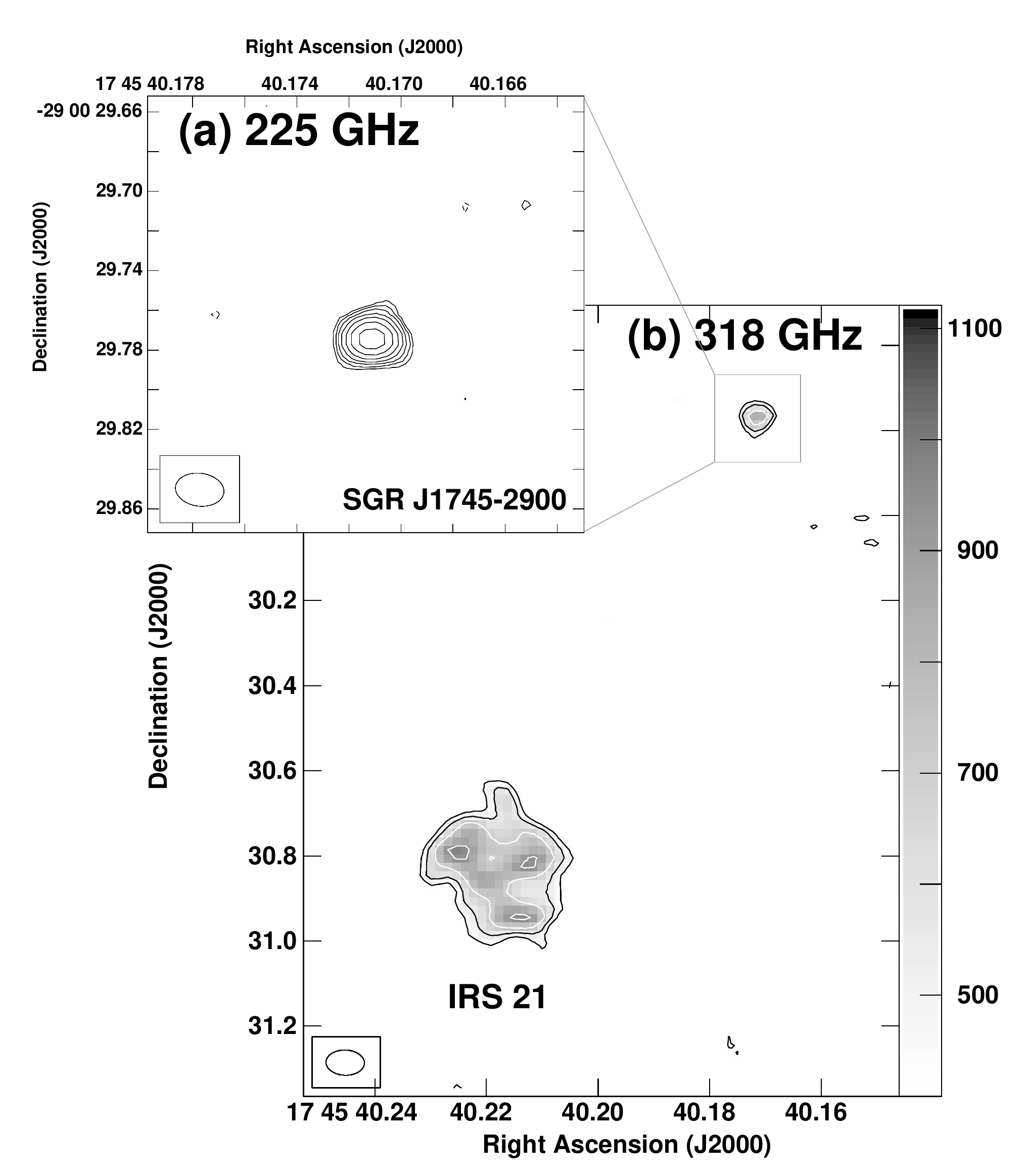}
\caption{(a) Contour image of the magnetar SGR J1745-2900 observed on 2017-10-6 using 
ALMA at 225.75 GHz with a FWHM beam of 0.024"$\times$0.017" (PA 82$^{\rm o}$).
Contours are $\sigma_{\rm im}\times$($-$5.7, 5.7, 8, 11.3, 16, 22.6, 32, 45.3 ), 
where $\sigma_{\rm im}=22\mu$Jy beam$^{-1}$.
(b) ALMA image of the magnetar SGR J1745-2900 (top-right: a point source) and IRS 21 
(bottom-left: a complex)  observed on 2019-10-14 at 318 GHz with a FWHM beam of 
0.087"$\times$0.059" (89.5$^{\rm o}$). Contours are $\sigma_{\rm im}\times$($-$6, 6, 8, 11, 15), 
where $\sigma_{\rm im}=55\mu$Jy beam$^{-1}$. The FWHM beams are illustrated at the 
bottom-left corner in each image. Numbers on the grey scale bar are in units of 
$\mu$Jy beam$^{-1}$.
}
\label{fig2}
\end{figure}

We also re-processed ALMA archival data (2015.1.01080.S) for observations at 343.49 GHz  
by \cite{tsu2017}. The observations were carried out at four epochs 2016-4-23, 2016-8-30, 
2016-8-31 and 2016-9-8, for durations of 3h, 1h, 2h and 3h, respectively. The 2016-4-23 
observation was in the C36-2/3 array configuration, and other three observations
were in the C40-6 configuration. The pipeline-calibrated datasets were obtained by 
executing the pipeline script ``{\bf scriptForPI.py}''. We then adjusted the input 
parameters in the CASA-Python script used for imaging the 225-GHz data, and applied 
the script to the 343-GHz data. The spatial resolution of the first 343-GHz image 
from epoch 2016-4-23 is relatively poor, with FWHM beam of 0.35"$\times$0.33" 
($-79^{\rm o}$), and the emission from the magnetar SGR J1745-2900 appears to be 
contaminated by surrounding extended emission. The final image of SGR J1745-2900 
was made by applying a high-pass baseline filter, so that only long-baseline data 
were included (${\rm >100~k\lambda}$), thus filtering out extended ($>2$") emission
features. An rms noise of 13 mJy beam$^{-1}$ was achieved for the epoch 2016-4-23, 
and the flux density of 3.29$\pm$0.34 mJy was determined for the magnetar at 343 GHz. 
The observations from the later epochs at 343 GHz have a typical angular resolution 
of 0.1". The rms noise values were 0.15, 0.05 and 0.12 mJy beam$^{-1}$ for the  
2017-8-30, 2017-8-31 and 2017-9-8 images, respectively. The magnetar SGR J1745-2900 
is detected in all four epochs at 343-GHz, and the flux densities are determined at 
a level of 10$\sigma$ or better using 2D-Gaussian fitting. The measurements are 
reported in Table 3, including a 5\% error in the flux-density calibration \citep{bon2018}.

A high-resolution (0.087"$\times$0.059", PA = 89.5$^{\rm o}$) ALMA observation was carried out 
at 320 GHz in the C43-6 array configuration on 2019-10-14 (2018.A.00052.S of P.I. Mark Morris).
The data was initially processed with the pipeline script ``{\bf scriptForPI.py}''. Subsequently, 
we made further corrections for the residual errors with the CASA-Python script described above 
and then imaged the region containing the GC magnetar, Sgr A*, and IRS 21 with the two lower-frequency 
sub-bands centered at 318 GHz, which have relatively stable phase and less contaminations 
from molecular lines in the circumnuclear disk (CND). An rms noise of 55 $\mu$Jy beam$^{-1}$ was 
achieved. The magnetar is significantly detected with a flux density of 1.32$\pm$0.15 mJy. 
The uncertainties include a 5\% error in the determination of the flux-density scale for 
ALMA observations. Fig. 2b shows the 318-GHz image of the region containing SGR J1745-2900 
and IRS 21.

\begin{figure*}[th!]
\centering
\includegraphics[angle=0,width=145mm]{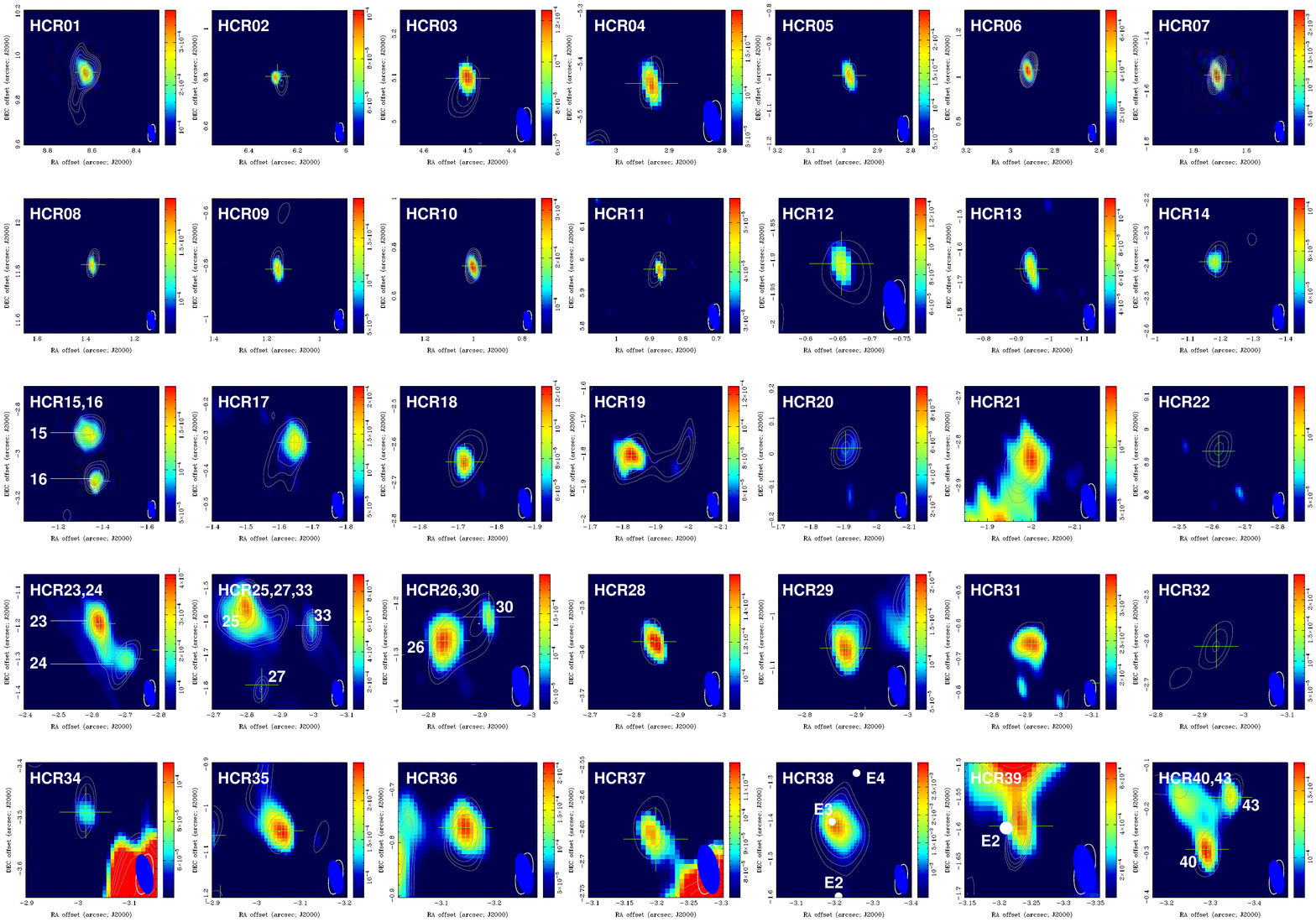} \\
\includegraphics[angle=0,width=145mm]{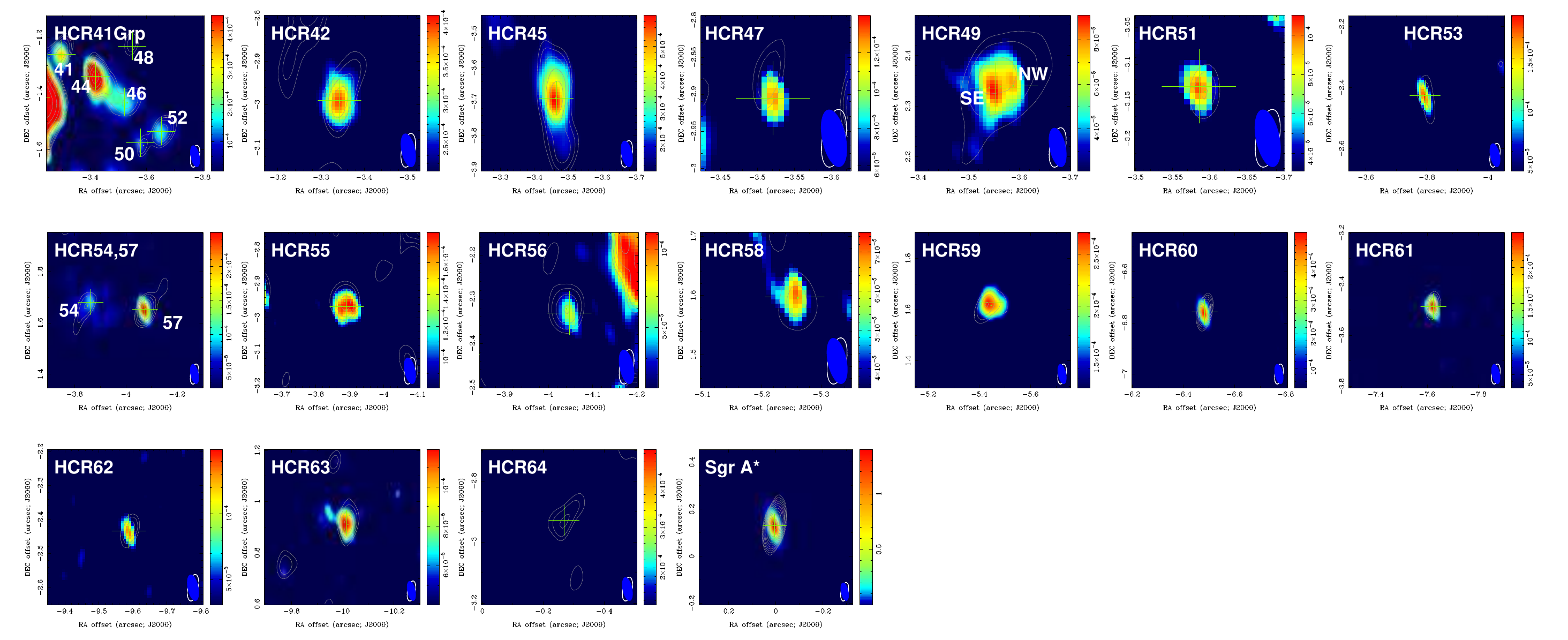}
\caption{Images of individual HCRs, color for 44.6 GHz intensity and contours for 33 GHz 
from an image made with a long baseline filter of $>500~{\rm k}\lambda$ except for HCR39 
which was made with a longer baseline filter $>1000~{\rm k}\lambda$ to separate the 
point-like source from the contamination of the extended emission associated with the 
radio core IRS 13E. The IR stars (E2, E3, and E4) detected in the IRS 13E region \citep{zhu2020}
are marked with white dots (see HCR38 and HCR39 panels). The color wedges scale the 
intensity of 44.6-GHz emission in units of Jy beam$^{-1}$. The contours are 
$\sigma_{\rm im} \times 2^{\rm n/2}$ where n starts from $\rm n_1$ until reaching the 
source peaks. The values of $\sigma_{\rm im}$ and $\rm n_1$ corresponding to the individual 
HCR image panels are listed in Column 11 of Table 2. Bottom-right corners are the FWHM 
beams at 44.6 and 33 GHz: 0.08"$\times$0.03" (PA = 12$^{\rm o}$) in blue and 0.08"$\times$0.03" 
(PA = -11$^{\rm o}$) in white, respectively.
}
\label{fig3}
\end{figure*}

\section{Hypercompact radio sources}

\subsection{Search procedure and selection criteria}

The angular resolutions of the  A-array  observations at Ka- and Q-bands, 0.05", 
are nearly two orders of magnitude smaller in beam area than the 5.5-GHz beam used 
in our previous study of the GCCRs \citep{zmg2020}. At such a high spatial resolution, 
most of the emission components in the Sgr A West HII region have been resolved out. 
The emission from the HII sources produced overwhelming confusion at 5.5 GHz, which 
was the main issue preventing us from unambiguously detecting GCCRs in Sgr A West, 
given the relatively low angular resolution of 0.5".  The Ka-band JVLA, with both 
its superior angular resolution and sensitivity ($\sim8~\mu$Jy beam$^{-1}$), is well 
suited for significant detection of point-like hyper-compact radio sources (HCRs) 
at a level of $>$0.1 mJy. Thus, we have an unprecedented opportunity for study of 
the compact radio sources associated with massive stellar remnants, such as pulsars, 
magnetars and accreting compact stellar remnants. Unlike the free-free emission of 
the HII components, the radio emission from such objects is expected to be nonthermal. 
Our search has therefore been focused on using the Ka- and Q-band images to identify 
nonthermal HCRs within the Galaxy's central parsec. We proceeded in three steps 
as follows: 

\vskip 5pt
\parindent 0pt
{
(1) We initially found approximately a thousand compact radio sources having angular 
size of $\lesssim 0.5$" based on a Ka-band image constructed including all baselines 
with a robust weighting parameter, R=0.25 \citep{bri1995}, and a FWHM beam of 
0.12"$\times$0.06" (PA = $-11^{\rm o}$). A sensitivity threshold of 
S$_{\rm 33\/GHz}$/$\sigma_{\rm rms}$ > 6, $\sigma_{\rm rms}\approx7~\mu$Jy beam$^{-1}$,
was applied in the initial search. Bright intensity maxima in the HII emission 
components might have been included in the compact-source sample, producing false 
detections for the nonthermal compact radio sources. To ensure that the ultimate 
sample contains only compact nonthermal radio sources, two further steps were  
carried out:
\vskip 5pt

(2) The Ka-band image was reconstructed with a robust weight R=$-$0.3 to 
down-weight the contribution from short baselines and also a high-pass
baseline filter ($>500$ k$\lambda$) was applied. So we can separate point-like 
sources from extended emission. The final cleaned image was convolved with a 
beam having FWHM = 0.08"$\times$0.03" (PA = $-11^{\rm o}$), similar to the 
synthesized beam of the Q-band observations, but keeping the position angle 
of the Ka-band synthesized beam determined by the uv-data sampling. We 
narrowed the list for those sources that are detected at Ka-band by keeping 
only those sources above a S$_{\rm 33}$/$\sigma_{\rm rms}=15$ threshold and 
limiting the size of the major axis to be less than 0.1 arcsec ($\theta_{\rm maj}<0.1$") 
based on a 2D Gaussian fitting. The number of candidates in the list was thereby 
reduced to $\sim$100. The remaining sub-sample only contains the hyper-compact 
members of the initial sample.

\vskip 5pt
(3) Finally, we used the high-resolution Q-band image, with $\sigma_{\rm rms}=17~\mu$Jy beam$^{-1}$,
FWHM = 0.08"$\times$0.03" (PA = 12$^{\rm o}$), to narrow down the list of HCRs
by imposing a conservative limit of $S_{44.6}/\sigma_{\rm im}>10$ on the detection significance 
and an upper limit on the positional offset between Ka- and Q-band positions less than $3\sigma$,
where $\sigma$ is given in Table 2 with a typical value of a few milli-arcsec (mas), based on 
2D-Gaussian fits at the two frequencies. We note that the locations of HII peaks depend 
on the uv-sampling. The uv data in the Q-band observations were primarily sampled in a 
positive hour-angle range while the Ka-band data were sampled in a negative hour-angle 
range. Consequently, the sources having a significant offset between the Q- and Ka-band 
images are suspected to be HII peaks and are therefore rejected from the HCR sample.  
}

\parindent=3.5mm
The equal beam sizes of the Q- and Ka-band images allow us to determine 
reliable spectral indices that will be used to further distinguish the source types. 
We do not use spectral indices as a selection parameter for the HCRs, given 
a wide range of values for the spectral indices covered by the radio sources 
associated with pulsars, magnetars and stellar-mass black holes. 
We note that the primary 15$\sigma$ significance cutoff used for 
Ka-band sources and the lower 10$\sigma$ cutoff for Q-band were chosen for
the derivation of significant spectral indices, given the poorer
sensitivity of the Q-band data. Neglecting fitting errors, for a weak point-like source,
we expect a maximum uncertainty of $\sigma_\alpha\sim0.4$ in spectral index.

\parindent=3.5mm
Those candidates having consistent results from the Gaussian fitting to the Ka- 
and Q-band images were included in the final sample of HCRs, consisting of 64 members.            
We note that the conservative search criteria for HCRs may miss highly variable  
and relatively weaker sources having a steep spectrum, given that the sensitivity of 
Q-band observations is a factor of 2 poorer than that of Ka-band data. For example, 
a hyper-compact radio source in the IRS 7SW region (see Fig. 1) was rejected from 
the HCR sample because its Ka-band flux density below the 15$\sigma$ threshold.  

\vskip 5pt
In short, from the JVLA A-array images observed at 33 and 44.6 GHz on 
2021-9-11 and 2021-9-11, we have identified 64 hyper-compact radio sources 
(HCRs) located inside Sgr A West within a radius of $13"$ from Sgr A* based on 
their compactness (a size of $\theta_{\rm maj}<0.1$") at a conservative 
significance level of $S/\sigma>15$ at 33 GHz and $S/\sigma>10$ at 44.6 GHz. 
Three exceptions which were not detected at 44.6 GHz on 2015-9-16 are also included:
HCR22, HCR32 and HCR64. HCR22 appears to be a strong candidate for the object that 
powers the X-ray PWN G359.95-0.04 (see Sec.4.1). HCR64 is a compact transient radio 
source associated with the microquasar discovered during the 2005 outburst of the 
X-ray transient XJ174540.0290031 \citep{mun2005,por2005,bow2005,zmga2009}; further 
discussion of this source is given in Sec.4.2. HCR32 appears to be a highly variable 
source. It was detected with a flux density of 0.5$\pm$0.1 mJy at 9 GHz at the epoch 
of 2014-4-17 but no significant detections were made in the 2019 and 2020 epochs 
at 9 GHz.  

One of the sources, HCR49, is associated with the high-velocity head-tail radio/IR 
source designated the "Bullet" \citep{yus1998,zg1998,zmga2009}; it now appears to
consist of at least two compact components.

\subsection{Images of hyper-compact radio sources}
Fig. 3 shows the images of all 64 HCRs in individual panels labelled with the HCR 
identification numbers, the colored images represent the 44.6-GHz HCRs and the 
superimposed contours are for the 33-GHz images made with baselines exceeding
$500{\rm k\lambda}$. For those panels containing more than one HCRs, each of 
the HCRs is labelled with its HCR ID. The green plus symbols mark the positions 
determined from the 44.6-GHz data. For HCR22, HCR32 and HCR64, which were
not detected at 44.6 GHz, the 33-GHz positions are shown. 

\subsection{A catalog for hyper-compact radio sources}

Table 2 lists 64 HCRs. Column 1 gives the ID numbers for the members of the  HCR catalog. 
Columns 2 and 3 are the J2000 coordinates, omitting the common part of 17$^{h}$45$^m$ 
in Right Ascension (RA) and $-29^{\rm o}$00' in Declination (Dec). With the exception
of HCR22, HCR32, and HCR64, the Q-band coordinates of the HCRs were determined with 
respect to Sgr A*'s equitorial coordinates at the J2000 epoch \citep{rei2014}. The 
positions of non-Q-band sources are determined using the Ka-band data. The 1$\sigma$ 
uncertainties in the last digits of RA and Dec. are given in parentheses. We note that, 
throughout this paper, the positions of the HCRs in the figures are labelled in the 
J2000 equatorial coordinate system with respect to the position of Sgr A*.
Columns 4 and 5 list the flux densities at 44.6 and 33 GHz with 1$\sigma$ uncertainties. 
Columns 6, 7 and 8 show the source sizes, deconvolved from the telescope beam, including 
major ($\theta_{maj}$) and  minor ($\theta_{min}$) axes, as well as the position angle (PA), 
all with their 1$\sigma$ uncertainties.
Column 9 lists the spectral index, $\alpha$ ($S\sim\nu^\alpha$), determined from the flux 
densities at 44.6 and 33 GHz. The parameters used to plot the contours in Fig. 3 for individual 
HCRs are listed in Column 10: $\sigma_{\rm im}$ representing the rms variations of the 
background regions near the sources and ${\rm n_1}$ being the integer corresponding to the 
multiplicative factor, $2^{0.5n}$, specifying the lowest contour.  Column 11 gives a brief 
note on individual HCRs. 

Finally, we examined the HCRs that can be spatially associated with 
the 7mm-IR(3.8$\mu$m) sources \citep{yus2014} by transfering their coordinates 
to the J2000 equatorial coordinate system used in this paper, in which Sgr A* is located at  
RA(J2000) = 17:45:40.0409, Dec(J2000) = $-$29:00:28.118 \citep{rei2014}. After corrections 
for the system errors (${\rm \Delta RA =0.0026^{\rm s}}$, ${\rm \Delta Dec = -0.049}$") 
that mainly caused by the difference of the  position of Sgr A* between \cite{rei2014} and 
\cite{yus2014}, 22 HCRs are identified with  possible 7mm-IR sources within an uncertainty
of 25 mas, the typical positional accuracy given for the 7mm-IR sources. The ID numbers of the 
possible 7mm-IR counterparts are given in Column 11 as well.   

\begin{figure}[!h]
\centering
\includegraphics[angle=0,width=89.5mm]{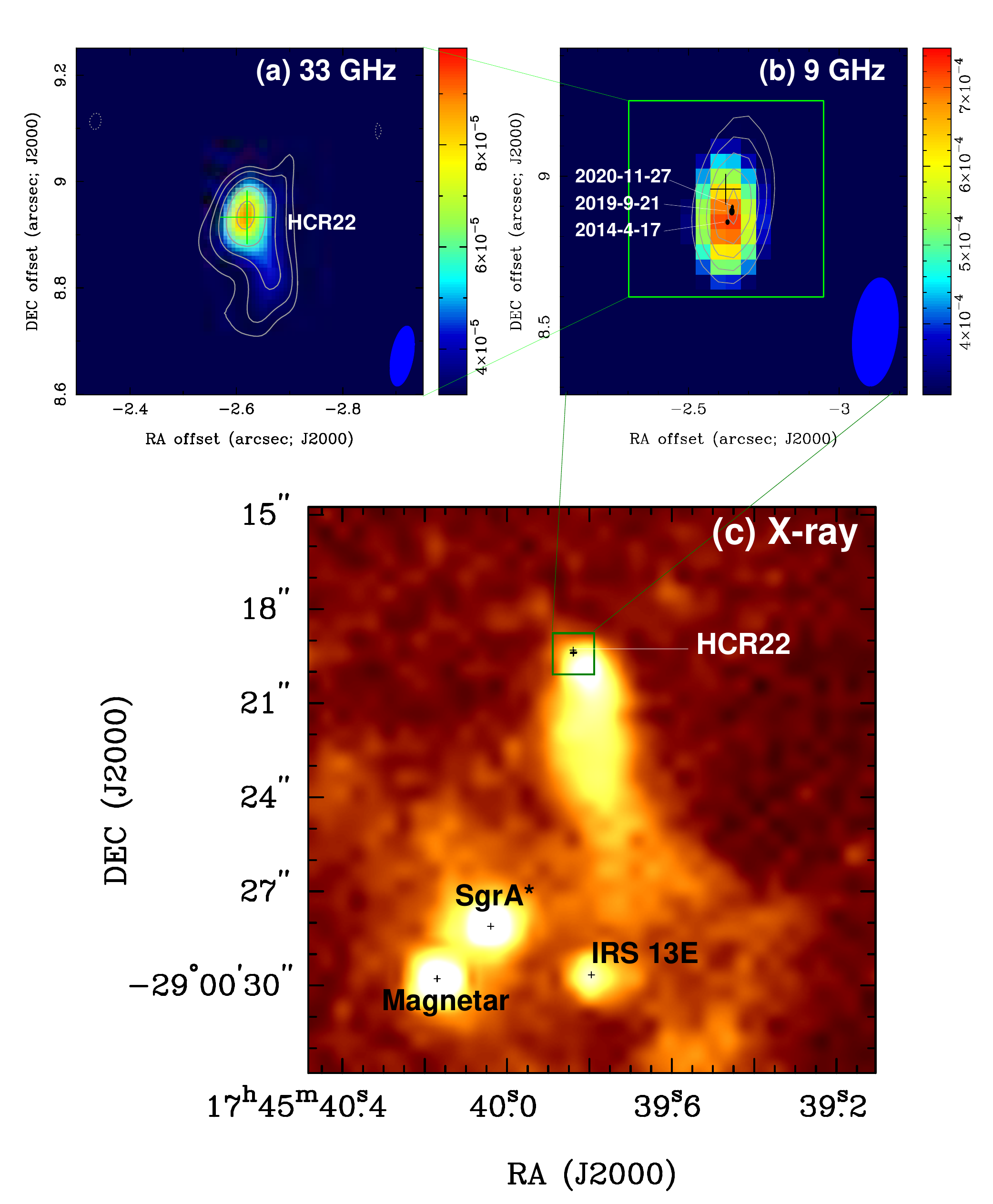}
\caption{
(a) The Ka-band image  made with the JVLA A-array data using all available baselines, 
showing the radio structure of HCR22, a candidate source of particles and energy powering the PWN.
The contours are 5 $\mu$Jy beam$^{-1}\times$ ($-4$, 4, 5.7, 8, 11.3, 16) with 
FWMH = 0.11"$\times$0.04", $-11^{\rm o}$ (bottom-right).
(b) JVLA 9-GHz images at epochs 2014-4-17 (color) and 2020-11-27 (contours). The contours are 
50 $\mu$Jy beam$^{-1}\times$ (64, 80, 96, 112, 128) with FWMH = 0.36"$\times$0.15", 
$-6^{\rm o}$ (bottom-right). The green plus marks the peak position at 33 GHz. 
The three black ellipses mark the peak positions of the counterpart at 9 GHz at 
the three observing epochs as labelled. The size of each ellipse scales the 
1 $\sigma$ uncertainty of the peak position determined from a 2D-Gaussian fitting. 
The color wedge scales the intensity in  units of Jy beam$^{-1}$ for the JVLA images.
(c) The Chandra X-ray image \citep[angular resolution = 0.5", from][]{zhu2018} of the 
PWN  G359.95-0.04 \citep{wan2006}. The green box at the tip of the PWN marks the region of HCR22 
as imaged at 9 GHz in panel b.
}
\label{fig4}
\end{figure}

\section{Typical cases for hyper-compact radio sources}
\subsection{HCR22 and the X-ray PWN G359.95-0.04}

\parindent 0pt
{\bf HCR22} is located at the northern termination of the extended X-ray source G359.95-0.04 
\citep{bag2003} that was suggested to be a pulsar wind nebula (PWN) \citep{wan2006, mun2008}. 
The source at 33 GHz can be characterized by a head (HCR22) with a tail extending $\sim0.3$" 
toward the south (Fig. 4a). The radio morphology is consistent with the X-ray structure, including 
the immediate angle of the tail, but with a much smaller scale. The radio emssion was also 
detected at 9 GHz in the three epoch images 2014-4-17 \citep{mzg2017}, 2019-9-21 and 2020-11-27 
(this paper) with a larger FWHM of 0.36"$\times$0.15" ($-6^{\rm o}$). At 9 GHz, the source shows no 
signficant variation in flux density, with measured flux densities of 1.70$\pm$0.09 mJy, 
1.53$\pm$0.06 mJy and 1.69$\pm$0.05 mJy at the three epochs, respectively. A spectral index 
of $-1.0\pm0.1$ is determined by the flux density of 0.43$\pm$0.03 mJy, integrated over the 
entire source, at 33 GHz and the mean flux density of 1.64$\pm$0.06 mJy averaged by the three 
epochs' data at 9 GHz. Assuming $\alpha_{33/44.6}\sim-1$,  a peak intensity of $<60~\mu$Jy beam$^{-1}$ 
at 44.6 GHz is extrapolated from the 33 GHz image (see Fig. 4). The non-detection of HCR22 
at Q-band is consistent with its steep spectrum. In addition, we noticed that the peak position 
of the source at 9 GHz moved towards north as time increases. From a least-squares fitting 
of the source position at the three epochs, we find a significant proper motion of the compact 
source at 9 GHz in the declination direction: giving 
$\mu_\alpha =-2.0\pm1.0$ mas y$^{-1}$ and $\mu_\delta =7.3\pm1.0$ mas y$^{-1}$
(Fig. 4a).  For a distance of 8 kpc to the Galactic center, this proper motion
corresponds to a velocity of $270\pm40$ km s$^{-1}$ projected onto the plane of the sky, 
which is consistent with the source's projected proximity to Sgr A*. The location, steep 
spectrum, head-tail structure and orientation of HCR22 at 33 GHz along with the significant 
northward motion of the 9-GHz peak, suggest that HCR22 is plausibly the candidate source of 
energetic particles that are responsible for the X-ray emission of PWN G359.95-0.04.

\parindent=3.5mm    
\subsection{Microquasar of X-ray transient CXOGC J174540.0-290031}
\parindent 0pt
{\bf HCR64} is the radio counterpart of the microquasar associated with the X-ray 
transient CXOGC J174540.0-290031 that  was discovered by the Chandra X-ray observatory 
\citep{mun2005,por2005}. The radio emission from this transient was found with the VLA 
\citep{bow2005,zmga2009}. Fig. 5 shows the 2015 Ka-band image (contours) overlaid on the 
color K-band image of 2005. A double source was detected  in the 2005 VLA observations 
at 22.5 GHz \citep{zmga2009} with a SW compact component (K40) associated with the core 
at the flux density of  3.4$\pm$0.1 mJy. The position offset between HCR64 (2015) and 
the SW component, K40, observed in 2005 appears to be insignificant, so we follow 
\citet{zmga2009} in identifying this as the core. The bright component 0.5" NE of the 
core in the 2005 image is  not detected in 2015. The NE component was 3.5$\pm$0.1 mJy 
during the 2005 outburst, which appeared to be launched from the microquasar
\citep{mun2005,por2005,bow2005,zmga2009}. The core (K40) was detected with a flux 
density of 1.1$\pm0.1$ mJy at 22.5 GHz at the early epochs in 1991 and 1999 
\citep{zmga2009}, representing quiescent level of the microquasar. 

\parindent=3.5mm

In our Ka-band image, which was observed 10 years after the 2005 outburst, the flux density 
of this compact radio source at 33 GHz was 0.60$\pm$0.07 mJy. The 33-GHz flux density appears to be consistent 
with that observed at 22.5 GHz during 1991 and 1995, when the microquasar was apparently in 
its quiescent state. From the ratio of the flux densities at 33 and 22.5 GHz, we determine 
a spectral index of $\alpha = -1.6\pm0.4$ assuming that the flux density in the quiescent 
state is not variable. The 3-$\sigma$ upper limit of 0.15 mJy at 44.6 GHz is consistent 
with this derived steep spectrum. 

\parindent=3.5mm
\begin{figure}[!h]
\centering
\includegraphics[angle=0,width=99.5mm]{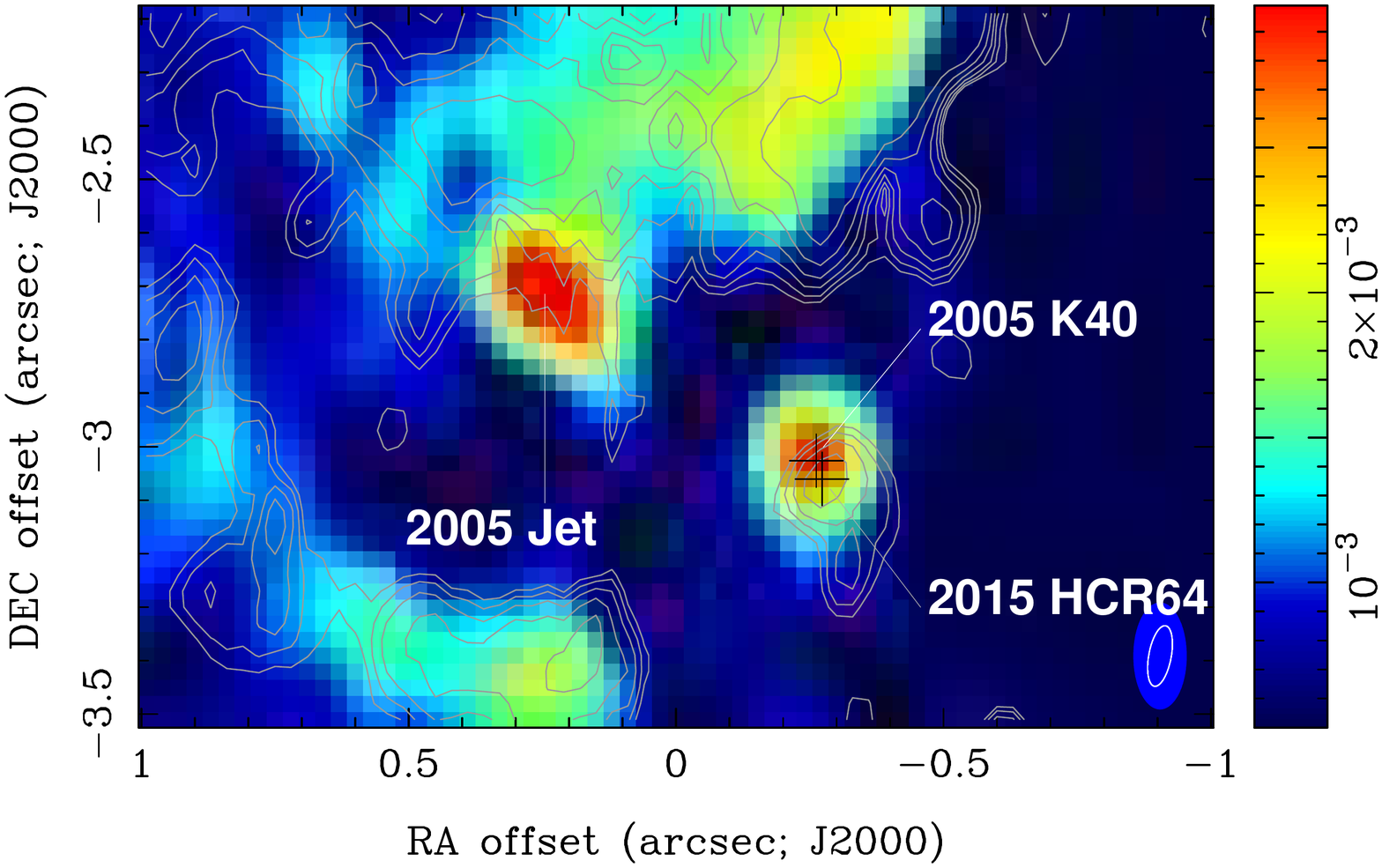}
\caption{Images of the radio emission from the micro-quasar correspoding to
the X-ray transient CXOGC J174540.0-290031, observed with the JVLA in the A-array 
at 33 GHz (contours) and 22.5 GHz (colors) on 2015-9-11 and 2005-9-16, respectively. 
At the bottom-right corner,  the FWHM beams  of 0.11"$\times$0.04" ($-11^{\rm o}$) and 
0.2"$\times$0.1" (0$^{\rm o}$) at 33 and 22.5 GHz are shown with a white and filled-blue ellipses.
Contour values are $\sigma_{\rm im}\times$ (5.7, 8, 11.3, 16, 22.6 and 32), 
where $\sigma_{\rm im}=7\mu$Jy beam$^{-1}$. The two plus symbols mark the positions 
of K40 at 22.5 GHz in 2005  and of HCR22  at 33 GHz in 2015.}
\label{fig5}
\end{figure}

\begin{table*}[ht!]
\scriptsize
\centering
\tablenum{2}
\setlength{\tabcolsep}{1.mm}
\caption{A catalog of hypercompact radio sources at 44.6 and 33 GHz}
\begin{tabular} {lcccccccrcc}
\hline\hline \\
{ID}&
{RA(J2000)}&
{Dec(J2000)}&
{$\left[S\pm\sigma\right]_{44.6}$}&
{$\left[S\pm\sigma\right]_{33}$}&
{$\theta_{maj}\pm\sigma$}&
{$\theta_{min}\pm\sigma$}&
{PA$\pm\sigma$}&
{$\alpha_{44.6/33}\pm\sigma$}&
{$\sigma_{\rm im}$, n$_1$}&
{Notes}\\
{} &
{(~~~$^{\rm s}$~~~)} &
{(~~~"~~~)}&
{(mJy)} &
{(mJy)} &
{(mas)} &
{(mas)} &
{(deg)} &
        &
{($\mu$Jy bm$^{-1}$)}\\
{(1)}&
{(2)}&
{(3)}&
{(4)}&
{(5)}&
{(6)}&
{(7)}&
{(8)}&
{(9)}&
{(10)}&
{(11)}\\
\hline\\
 HCR01 &40.6981(3) &18.323( 6) &  $0.95\pm0.11 $& $1.00\pm0.16 $& $ 74.\pm 19. $& $ 56.\pm 17. $& $ 36.\pm23.$& $-0.17\pm0.65$&7.0, ~6&{ID9$^{25}$}\\
 HCR02 &40.5194(4) &27.445( 7) &  $0.20\pm0.02 $& $0.15\pm0.02 $& $ 68.\pm 11. $& $ 26.\pm 30. $& $ 30.\pm45.$& $ 0.96\pm0.60$&7.0, ~5&\\
 HCR03 &40.3836(1) &23.150( 3) &  $0.27\pm0.02 $& $0.20\pm0.02 $& $ 57.\pm 14. $& $ 40.\pm  8. $& $150.\pm36.$& $ 1.00\pm0.48$&7.0, ~5&\\
 HCR04 &40.2641(1) &33.686( 3) &  $0.26\pm0.02 $& $0.22\pm0.01 $& $ 45.\pm 13. $& $ 16.\pm  8. $& $100.\pm13.$& $ 0.55\pm0.38$&7.0, ~7&IRS 9W$^{21}$  \\
 HCR05 &40.2679(1) &29.243( 3) &  $0.22\pm0.02 $& $0.19\pm0.02 $& \multicolumn{3}{c}{unresolved}& $ 0.49\pm0.52$&7.0, ~8&{ID4$^{25}$}  \\
 HCR06 &40.2629(1) &27.217( 1) &  $0.80\pm0.04 $& $0.75\pm0.04 $& $ 20.\pm  6. $& $ 15.\pm  9. $& $ 80.\pm42.$& $ 0.21\pm0.25$&7.0, ~5&{ID2$^{25}$}\\
 HCR07 &40.1708(1) &29.788( 1) &  $2.54\pm0.13 $& $1.90\pm0.10 $& $ 20.\pm  8. $& $  8.\pm  8. $& $ 56.\pm47.$& $ 0.96\pm0.24$&7.0, ~9&Magnetar$^1$\\
 HCR08 &40.1449(4) &16.419(13) &  $0.23\pm0.02 $& $0.21\pm0.02 $& $ 19.\pm 10. $& $ 13.\pm 20. $& $ 80.\pm24.$& $ 0.30\pm0.32$&7.0, ~6&\\
 HCR09 &40.1286(1) &29.060( 4) &  $0.21\pm0.02 $& $0.27\pm0.02 $& $ 49.\pm 23. $& $ 18.\pm 11. $& $ 84.\pm30.$& $-0.83\pm0.46$&7.0, ~7&{ID3$^{25}$}\\
 HCR10 &40.1166(1) &27.526( 1) &  $0.36\pm0.03 $& $0.33\pm0.03 $& $ 15.\pm 16. $& $  4.\pm 17. $& $ 22.\pm44.$& $ 0.29\pm0.36$&7.0, ~5&{ID1$^{25}$}\\
 HCR11 &40.1067(6) &22.277( 9) &  $0.11\pm0.02 $& $0.17\pm0.01 $& $ 61.\pm 31. $& $ 26.\pm  9. $& $ 82.\pm21.$& $-1.45\pm0.68$&7.0, ~5&IRS 7E$^2$  \\
 HCR12 &39.9905(6) &30.154(12) &  $0.19\pm0.01 $& $0.33\pm0.03 $& $ 51.\pm 10. $& $ 44.\pm  6. $& $ 38.\pm18.$& $-1.83\pm0.42$&7.0, ~7&\\
 HCR13 &39.9683(6) &29.907(12) &  $0.25\pm0.03 $& $0.40\pm0.04 $& $ 95.\pm 18. $& $ 34.\pm  6. $& $ 94.\pm28.$& $-1.56\pm0.57$&8.5, ~7&\\
 HCR14 &39.9502(1) &30.636( 3) &  $0.31\pm0.02 $& $0.40\pm0.04 $& $ 74.\pm 10. $& $ 59.\pm  9. $& $158.\pm64.$& $-0.85\pm0.36$&7.0, ~6&\\
 HCR15 &39.9380(3) &31.162( 6) &  $0.16\pm0.02 $& $0.29\pm0.02 $& $ 72.\pm 16. $& $ 18.\pm 33. $& $ 90.\pm20.$& $-1.97\pm0.53$&8.0, ~6&Mini-cavity$^7$\\
 HCR16 &39.9350(3) &31.370( 6) &  $0.59\pm0.07 $& $0.82\pm0.05 $& $ 81.\pm 17. $& $ 62.\pm 18. $& $134.\pm32.$& $-1.09\pm0.43$&8.0, ~6&Mini-cavity$^7$\\
 HCR17 &39.9151(3) &28.563( 7) &  $0.52\pm0.07 $& $0.69\pm0.11 $& $ 81.\pm 29. $& $ 60.\pm 21. $& $166.\pm90$& $-0.94\pm0.66$&7.0, ~6&$\epsilon^8$\\
 HCR18 &39.9096(1) &30.895( 4) &  $0.17\pm0.02 $& $0.31\pm0.03 $& $ 48.\pm 14. $& $ 29.\pm 20. $& $134.\pm24.$& $-1.99\pm0.50$&8.0, ~6&Mini-cavity$^7$  \\
 HCR19 &39.9013(1) &30.053( 3) &  $0.41\pm0.03 $& $0.54\pm0.04 $& $ 82.\pm  9. $& $ 60.\pm 11. $& $ 98.\pm24.$& $-0.91\pm0.34$&8.0, ~6&Mini-cavity$^9$, {ID18$^{25}$} \\ 
 HCR20 &39.8952(6) &28.230(12) &  $0.20\pm0.03 $& $0.38\pm0.04 $& $ 60.\pm 23. $& $ 56.\pm 23. $& $144.\pm59.$& $-2.13\pm0.61$&7.0, ~6&$\epsilon^8$  \\
 HCR21 &39.8883(3) &31.088( 7) &  $0.32\pm0.04 $& $0.56\pm0.03 $& $ 87.\pm 20. $& $ 64.\pm 16. $& $158.\pm77.$& $-1.86\pm0.49$&8.0, ~6&  \\
 HCR22 &39.8408(3) &19.316( 6) &  $<0.06$       & $0.15\pm0.02 $& $ 73.\pm 16. $& $ 50.\pm 61. $& $ 73.\pm82.$& N/A&7.0, ~6&PWN$^3$  \\
 HCR23 &39.8407(1) &29.442( 4) &  $0.92\pm0.08 $& $1.00\pm0.11 $& $ 84.\pm 12. $& $ 44.\pm 30. $& $ 14.\pm 8.$& $-0.28\pm0.46$&8.0, ~7&Mini-cavity$^{10}$, {ID20$^{25}$} \\
 HCR24 &39.8345(1) &29.548( 3) &  $0.19\pm0.01 $& $0.20\pm0.01 $& \multicolumn{3}{c}{unresolved}& $-0.17\pm0.34$&8.0, ~7&Mini-cavity$^{10}$, {ID19$^{25}$}\\
 HCR25 &39.8272(1) &29.810( 4) &  $0.88\pm0.08 $& $1.29\pm0.13 $& $ 49.\pm 18. $& $ 34.\pm 19. $& $ 20.\pm52.$& $-1.27\pm0.45$&8.0, ~7&IRS 13E-SE$^6$, {ID32$^{25}$} \\
 HCR26 &39.8247(1) &29.521( 3) &  $0.29\pm0.02 $& $0.25\pm0.02 $& \multicolumn{3}{c}{unresolved}& $ 0.49\pm0.42$&8.0, ~7&IRS 13E-NE$^{11}$, {ID21$^{25}$}   \\
 HCR27 &39.8236(3) &30.039( 7) &  $0.21\pm0.03 $& $0.37\pm0.04 $& $ 45.\pm 22. $& $ 25.\pm 28. $& $133.\pm52.$& $-1.88\pm0.64$&8.0, ~7&IRS 13E-SE$^6$  \\
 HCR28 &39.8235(3) &31.825( 6) &  $0.35\pm0.04 $& $0.45\pm0.03 $& $ 73.\pm 16. $& $ 36.\pm  8. $& $ 20.\pm16.$& $-0.83\pm0.47$&7.0, ~8&Mini-cavity$^{12}$  \\
 HCR29 &39.8210(1) &29.311( 4) &  $0.30\pm0.03 $& $0.50\pm0.03 $& $ 44.\pm 12. $& $ 34.\pm 20. $& $148.\pm36.$& $-1.70\pm0.41$&8.0, ~7&IRS 13E-NE$^{11}$, {ID23$^{25}$}  \\
 HCR30 &39.8181(1) &29.475( 4) &  $0.20\pm0.02 $& $0.28\pm0.02 $& $ 58.\pm  5. $& $ 22.\pm  2. $& $164.\pm 3.$& $-1.12\pm0.47$&8.0, ~7&IRS 13E-NE$^{11}$, {ID22$^{25}$}  \\
 HCR31 &39.8180(1) &28.902( 6) &  $0.63\pm0.03 $& $0.93\pm0.06 $& $ 87.\pm  5. $& $ 42.\pm  7. $& $ 98.\pm 7.$& $-1.29\pm0.26$&8.0, ~9&IRS 13E-NE$^{13}$ \\
 HCR32 &39.8165(2) &30.861( 7) &  $<0.06       $& $0.42\pm0.03 $& $ 40.\pm  5. $& $ 11.\pm 22. $& $ 72.\pm 4.$& N/A&7.0, ~7&Mini-cavity$^{14}$  \\
 HCR33 &39.8121(1) &29.860( 3) &  $0.35\pm0.03 $& $0.58\pm0.05 $& $ 48.\pm 14. $& $ 16.\pm 10. $& $ 14.\pm22.$& $-1.68\pm0.43$&8.0, ~7&IRS 13E-SE$^6$, {ID30$^{25}$}   \\
 HCR34 &39.8107(1) &31.739( 4) &  $0.30\pm0.03 $& $0.45\pm0.04 $& $ 93.\pm 14. $& $ 45.\pm  6. $& $  8.\pm12.$& $-1.35\pm0.46$&7.0, ~6&Mini-cavity$^{12}$  \\
 HCR35 &39.8075(1) &29.300( 3) &  $0.42\pm0.03 $& $0.43\pm0.05 $& $ 55.\pm 10. $& $ 22.\pm 26. $& $ 62.\pm24.$& $-0.08\pm0.42$&7.0, ~8&IRS 13E-NE$^{11}$, {ID24$^{25}$}  \\
 HCR36 &39.8007(3) &29.020( 6) &  $0.31\pm0.04 $& $0.58\pm0.04 $& $ 42.\pm  8. $& $ 27.\pm 16. $& $ 56.\pm27.$& $-2.08\pm0.50$&8.0, ~6&IRS 13E-N$^{15}$, {ID28$^{25}$}   \\
 HCR37 &39.7967(6) &30.914( 9) &  $0.18\pm0.04 $& $0.41\pm0.03 $& $ 72.\pm 25. $& $ 41.\pm 55. $& $ 36.\pm44.$& $-2.73\pm0.79$&8.0, ~7&IRS 2-N$^{16}$  \\
 HCR38 &39.7961(1) &29.662( 3) &  $5.50\pm0.38 $& $3.35\pm0.42 $& $ 78.\pm 18. $& $ 42.\pm 18. $& $ 52.\pm21.$& $ 1.65\pm0.47$&8.0, ~12&IRS 13E$^5$, {ID33$^{25}$}   \\
 HCR39 &39.7941(3) &29.844( 9) &  $1.27\pm0.12 $& $1.14\pm0.09 $& $ 76.\pm 17. $& $ 29.\pm  5. $& $ 90.\pm 7.$& $ 0.36\pm0.41$&8.0, ~6&IRS 13E-S$^{17}$, {ID34$^{25}$}  \\
 HCR40 &39.7895(1) &28.540( 4) &  $0.58\pm0.05 $& $0.69\pm0.04 $& $ 94.\pm  9. $& $ 54.\pm  3. $& $ 91.\pm 5.$& $-0.58\pm0.34$&8.0, ~6&Triplet$^{18}$ \\
 HCR41 &39.7891(3) &29.508( 7) &  $0.90\pm0.12 $& $1.10\pm0.07 $& $ 61.\pm 20. $& $ 41.\pm 30. $& $150.\pm38.$& $-0.67\pm0.49$&8.0, ~7&IRS 13E-W$^{19}$  \\
 HCR42 &39.7857(3) &31.239( 6) &  $0.67\pm0.05 $& $1.00\pm0.06 $& $ 84.\pm 11. $& $ 52.\pm  7. $& $166.\pm12.$& $-1.33\pm0.32$&8.0, ~7&IRS 2  \\
 HCR43 &39.7855(1) &28.421( 4) &  $0.30\pm0.03 $& $0.48\pm0.05 $& $ 68.\pm 16. $& $ 46.\pm 11. $& $161.\pm30.$& $-1.56\pm0.49$&8.0, ~6&Triplet$^{18}$  \\
 HCR44 &39.7799(1) &29.587( 4) &  $1.23\pm0.13 $& $1.23\pm0.10 $& $ 46.\pm 18. $& $ 22.\pm 17. $& $ 36.\pm33.$& $ 0.00\pm0.45$&8.0, ~7&IRS 13E-W$^{19}$, {ID35$^{25}$}   \\
 HCR45 &39.7764(3) &31.938( 3) &  $1.48\pm0.09 $& $1.15\pm0.08 $& $ 96.\pm 10. $& $ 40.\pm  4. $& $  1.\pm 5.$& $ 0.84\pm0.31$&8.0, ~8&IRS 2L, {ID36$^{25}$}  \\
 HCR46 &39.7721(1) &29.678( 7) &  $0.64\pm0.06 $& $0.68\pm0.04 $& $ 90.\pm 11. $& $ 56.\pm 14. $& $ 78.\pm16.$& $-0.20\pm0.36$&8.0, ~7&IRS 13E-W$^{19}$  \\
 HCR47 &39.7721(2) &31.154( 9) &  $0.18\pm0.03 $& $0.15\pm0.03 $& $ 44.\pm 18. $& $ 21.\pm  6. $& $ 82.\pm10.$& $ 0.61\pm0.90$&7.0, ~5&IRS 2  \\
 HCR48 &39.7699(1) &29.479( 7) &  $0.10\pm0.02 $& $0.19\pm0.01 $& $ 35.\pm 24. $& $ 16.\pm 14. $& $ 59.\pm34.$& $-2.13\pm0.73$&8.0, ~7&IRS 13E-W$^{19}$ \\
 HCR49 &39.7697(3) &25.913( 7) &  $0.17\pm0.02 $& $0.21\pm0.04 $& $ 65.\pm 19. $& $ 39.\pm 13. $& $ 78.\pm14.$& $-0.70\pm0.78$&7.0, ~7&Bullet SE$^4$\\
 \dots &39.7671(4) &25.909( 6) &  $0.13\pm0.02 $& $0.16\pm0.04 $& $ 63.\pm 14. $& $ 23.\pm 05. $& $ 5.\pm10.$& $-0.69\pm1.00$&\dots&Bullet NW$^4$\\
 HCR50 &39.7677(1) &29.824( 3) &  $0.27\pm0.02 $& $0.30\pm0.03 $& $ 47.\pm 13. $& $ 28.\pm 14. $& $ 72.\pm31.$& $-0.35\pm0.42$&8.0, ~7&IRS 13E-W$^{19}$  \\
 HCR51 &39.7672(1) &31.378( 3) &  $0.19\pm0.01 $& $0.24\pm0.02 $& \multicolumn{3}{c}{unresolved}& $-0.78\pm0.40$&7.0, ~7&IRS 2-W  \\
 HCR52 &39.7620(4) &29.783( 4) &  $0.31\pm0.04 $& $0.47\pm0.03 $& $ 24.\pm  1. $& $ 16.\pm  1. $& $ 60.\pm 3.$& $-1.38\pm0.49$&8.0, ~7&IRS 13E-W$^{19}$ \\
 HCR53 &39.7513(1) &30.675( 3) &  $0.36\pm0.02 $& $0.30\pm0.03 $& $ 80.\pm 11. $& $ 14.\pm 41. $& $ 20.\pm 3.$& $ 0.61\pm0.34$&7.0, ~9&IRS 13E-SW$^{20}$ \\
 HCR54 &39.7461(3) &26.567(10) &  $0.11\pm0.02 $& $0.15\pm0.02 $& $ 28.\pm 10. $& $ 12.\pm 21. $& $ 60.\pm10.$& $-1.03\pm0.78$&7.0, ~8&IRS 34SW$^{22}$  \\
 HCR55 &39.7439(4) &31.215( 6) &  $0.57\pm0.08 $& $0.86\pm0.14 $& $ 81.\pm 22. $& $ 53.\pm 46. $& $108.\pm98.$& $-1.37\pm0.69$&7.0, ~7&  \\
 HCR56 &39.7320(1) &30.580( 4) &  $0.19\pm0.02 $& $0.36\pm0.02 $& $ 58.\pm  7. $& $ 38.\pm  6. $& $ 55.\pm10.$& $-2.12\pm0.43$&7.0, ~8&  \\
 HCR57 &39.7299(1) &26.595( 3) &  $0.33\pm0.03 $& $0.35\pm0.03 $& $ 38.\pm 17. $& $ 18.\pm 26. $& $ 60.\pm42.$& $-0.20\pm0.48$&7.0, ~8&IRS 34SW$^{22}$, {ID6$^{25}$}   \\
 HCR58 &39.6399(1) &26.651( 4) &  $0.19\pm0.02 $& $0.20\pm0.02 $& $ 76.\pm 18. $& $ 44.\pm 46. $& $124.\pm53.$& $-0.17\pm0.54$&7.0, ~8&IRS 6E$^{23}$   \\ 
 HCR59 &39.6255(3) &26.623( 4) &  $1.21\pm0.08 $& $1.40\pm0.17 $& $ 94.\pm 22. $& $ 68.\pm  7. $& $132.\pm35.$& $-0.48\pm0.47$&8.0, ~8&IRS 6E$^{23}$, {ID37$^{25}$}  \\
 HCR60 &39.5463(1) &35.005( 1) &  $0.60\pm0.04 $& $0.50\pm0.03 $& $ 31.\pm 12. $& $ 16.\pm 50. $& $ 90.\pm23.$& $ 0.61\pm0.30$&7.0, ~8&{ID7$^{25}$} \\
 HCR61 &39.4593(1) &31.736( 4) &  $0.34\pm0.03 $& $0.31\pm0.03 $& \multicolumn{3}{c}{unresolved}& $ 0.31\pm0.43$&7.0, ~9&{ID8$^{25}$} \\
 HCR62 &39.3096(1) &30.683( 1) &  $0.22\pm0.01 $& $0.18\pm0.01 $& $ 30.\pm  5. $& $  9.\pm  2. $& $ 10.\pm 2.$& $ 0.67\pm0.33$&7.0, ~8&  \\ 
 HCR63 &39.2771(4) &27.332(12) &  $0.22\pm0.04 $& $0.27\pm0.02 $& $ 51.\pm 33. $& $ 36.\pm 34. $& $ 42.\pm48.$& $-0.68\pm0.69$&7.0, ~8&  \\
 HCR64 &40.0201(9) &31.178(20) &  $<0.15$           & $0.60\pm0.07 $& $140.\pm 20. $& $ 70.\pm 10. $& $134.\pm10.$&N/A            &8.0, ~5  &micro-qso,K40$^{24}$\\
\\
\hline
\end{tabular}
\begin{tabular}{p{0.95\textwidth}}
{
\tiny
$^1$The radio counterpart of SGR J1745-2900 \citep{eat2013a}, the GC magnetar, and see Sec.4.3 of this paper. 
$^2$Located east of the IRS 7 bow-shock feature \citep{yus1991}.
$^3$A candidate cannonball of the PWN G359.95-0.04 \citep{wan2006}, and see Sec.4.1 of this paper.
$^4$The head of the bullet \citep{zmga2009},  resolved into two components. 
$^5$The radio core associated with IRS 13E. 
$^6$Located $\sim$0.5" SE of IRS 13E core \citep[e.g.,][]{tsu2017}.
$^7$Located within the mini-cavity.
$^8$The $\epsilon$ source that has been resolved into three components \citep{yus1990,zha1991}, S (HCR17), NE and NW (HCR20);
the three components correspond to RS6, RS5 and RS7 detected at 34.5 GHz \citep{yus2016}.
$^9$Located in the northern mini-cavity. 
$^{10}$Located at the northwestern rim of the mini-cavity. 
$^{11}$Located 0.5" NE of the IRS 13E radio core.
$^{12}$Located at the southwestern rim of the mini-cavity. 
$^{13}$Located 1" NE of the IRS 13E radio core.
$^{14}$Located at the western rim of the mini-cavity.
$^{15}$Located 0.7" N of the IRS 13E radio core.
$^{16}$Located N of the IRS 2.
$^{17}$Located 0.2" S of the IRS 13E radio core.
$^{18}$A triplet located 1.2" N of the IRS 13E radio core, consisting of three components S(HCR40), NW(HCR43) and NE with a size larger than the HCR upper limit. 
$^{19}$Located 0.3" W of the IRS 13E radio core, the HCR41 group consisting of six members of HCR41, 44, 46, 48, 50 and 52.
$^{20}$Located 1" SW of the IRS 13E radio core.
$^{21}$Located in the IRS 9W region.
$^{22}$Located in the IRS 34SW region.
$^{23}$Located in the IRS 6E region.
$^{24}$Located in the bar, a radio counterpart of the X-ray transients CXOGC J174540.029003 \citep{mun2005,bow2005,zmga2009} and see Sec.4.2 of this paper.
{$^{25}$The 7mm-sources are identified with IR stars emitting at 3.8 $\mu$m, among which
HCR05 (ID4), HCR06 (ID2), HCR10 (ID1), HCR57 (ID6), HCR60 (ID7) and HCR61 (ID8) are associated with
strong stellar winds \citep{yus2014}.}
}
\end{tabular}
\end{table*}

\subsection{The GC magnetar: SGR J1745-2900}  
\parindent 0pt
{\bf HCR07} is  one of the most recognizable HCR members, as it is associated with 
the GC magnetar and has a high-energy counterpart, SGR J1745-2900. The soft gamma-ray 
repeater was discovered by Swift during a large X-ray outburst on 2013 April 24 (MJD 56406), 
powered by a magnetar close to Sgr A*\citep{ken2013}. The magnetar hypothesis was further 
supported by NuSTAR detections of  periodic pulsed signal at 3.76 s \citep{mor2013}. 
With the observations by Chandra and Swift, \cite{rea2013} pinpointed the location of 
the magnetar at a projected distance of 2.4$\pm$0.3 arcsec from Sgr A*; and the authors 
also determined the source spin period and its derivative with high precision ($P=$3.7635537(2) s 
and $\dot{P} = 6.61(4) \times 10^{-12}$ s s$^{-1}$). The magnetar, SGR J1745-2900, 
was monitored by the Chandra X-ray observatory for six years  following the X-ray outburst 
in 2013 April, showing the long-term properties of the outburst  \citep{rea2020}.   

\parindent=3.5mm
Radio pulses from SGR J1745-2900 were first detected by the Effelsberg radio telescope and 
confirmed with other telescopes, including the Nancay telescope, JVLA, Jodrell Bank Observatory
\citep{eat2013a}, and the Australia Telescope Compact Array (ATCA) \citep{sha2013} at various 
frequencies between 1.5 and 19 GHz. High-frequency pulses were detected at 87, 101, 138, 154, 209 
and 225 GHz with the IRAM-30m telescope during the period between 2014-7-21 (MJD 56859) and 
2014-7-24 (MJD 56862) \citep{tor2015}. In their follow-up campaign, \cite{tor2017} detected 
high-frequency pulses with the IRAM-30m up to 291 GHz in the interval between 2015-3-4 (MJD 57085) 
and 2015-3-9 (MJD 57090). The high-frequency pulses were also detected  at 45 GHz with the 
Green Bank Telescope (GBT) on 2014-4-10 (MJD 56757) \citep{gel2017}.

\subsubsection{A collection of  flux-densitiy measurements from JVLA and ALMA}
In addition to the JVLA and ALMA flux densities determined from the images reported in this work, 
we also collected the data from the prior published literature. Table 3 assembles all the available 
data from JVLA and ALMA observations at radio wavelengths. The table is configured into two main 
column sections, each of which consists of six sub-columns: 
(1) observing date; 
(2) the corresponding modified Julian day (MJD);
(3) band center frequency in units of GHz;
(4) flux densities in units of mJy and 
(5) the corresponding 1$\sigma$ uncertainties. 
For the non-detections, a 3$\sigma$ value is given for the upper limits. Finally, 
alphabetical codes are designated for the  relevant references that are listed at the 
bottom notes of the table. The measurements reported for the first time in this paper 
are highlighted with a bold font.

\begin{table*}[!h]
\scriptsize
\centering
\tablenum{3}
\setlength{\tabcolsep}{1.7mm}
\caption{JVLA \& ALMA flux densities of SGR J1745-2900}
\begin{tabular}{lcccccclccccc}
\hline\hline \\
{Obs. date} &
{MJD}       &
{$\nu_0$ (GHz)}&
{S$_\nu$ (mJy)}&
{$\sigma$ (mJy)}&
{Ref.}& &
{Obs. date} &
{MJD}       &
{$\nu_0$ (GHz)}&
{S$_\nu$ (mJy)}&
{$\sigma$ (mJy)}&
{Ref.}\\
(1)&(2)&(3)&(4)&(5)&(6)&~~~~~~~~~~&
(1)&(2)&(3)&(4)&(5)&(6)\\
\hline \\
2011-08-04& 55776& 42   &  $<0.24^\dagger$  &\dots&a&&
2012-10-14& 56214& 21.2 &  $<0.60^\dagger$  &\dots&a\\
2012-10-14& 56214& 32   &  $<0.81^\dagger$  &\dots&a&&
2012-10-14& 56214& 41   &  $<0.78^\dagger$  &\dots&a\\
2012-12-22& 56283& 21.2 &  $<0.90^\dagger$  &\dots&a&&
2012-12-22& 56283& 32   &  $<0.94^\dagger$  &\dots&a\\
2012-12-22& 56283& 41   &  $<0.69^\dagger$  &\dots&a&&
2013-05-10& 56422&9     & 0.56              &0.011&g\\
2013-06-01& 56444&9     & 0.76  &0.015&g&&
2013-06-30& 56473&15    & 0.58  &0.012&g\\   
2013-07-13& 56486&9     &1 .47  &0.030&g&&
2013-10-26& 56591& 21.2 &  $<1.95^\dagger$  &\dots&a\\
2013-10-26& 56591& 32   &  $<1.45^\dagger$  &\dots&a&&
2013-10-26& 56591& 41   &  $<0.82^\dagger$  &\dots&a\\
2013-10-26& 56591& 41   &  0.7 & 0.4&e&&
2013-11-29& 56626& 21.2 &  $<2.04^\dagger$  &\dots&a\\
2013-11-29& 56626& 32   &  $<0.82^\dagger$  &\dots&a&&
2013-11-29& 56626& 41   &  $<0.70^\dagger$  &\dots&a\\
2013-11-29& 56626& 41   &  0.89 &    0.08&e&&
2013-12-29& 56656& 21.2 &  $<1.70^\dagger$  &\dots&a\\
2013-12-29& 56656& 32   &  $<0.86^\dagger$  &\dots&a&&
2013-12-29& 56656& 41   &  $<1.52^\dagger$  &\dots&a\\
2013-12-29& 56656& 41   &  1.20             &0.7&e&&
2014-01-01& 56658&15    & 2.09 & 0.042&g\\
2014-01-01& 56658&9      &1.18  &0.024&g&&
2014-02-15& 56703& 21.2 &  0.84 & 0.33&a\\
2014-02-15& 56703& 32   &  1.83&  0.10&a&&
2014-02-15& 56703& 41   &  1.85 & 0.07&a\\
2014-02-15& 56703& 41   &  2.1 &0.4&e&&
2014-02-21& 56709& 44.6 &  1.62&  0.04&a\\
2014-02-22& 56710&15     &1.07  &0.022&g&&
2014-02-22& 56710&9      &0.94 & 0.019&g\\
2014-03-09& 56731& 34.5 &  1.30&  0.01&a&&
2014-03-22& 56738& 21.2 &  2.79&  0.19&a\\
2014-03-22& 56738& 32   &  2.64&  0.05&a&&
2014-03-22& 56738& 41   &  1.24&  0.02&a\\
2014-03-22& 56738& 41   &  2.1& 0.3&e&&
2014-04-03& 56750&23    & 0.92 & 0.019&g\\
2014-04-03& 56750&43    & 0.54 & 0.011&g&&
2014-04-25& 56772&9     & 1.00 & 0.020&g\\
2014-04-25& 56772&15    & 1.22 & 0.025&g&&
2014-04-26& 56743& 21.2 &  0.90&  0.14&a\\
2014-04-26& 56743& 32   &  0.62&  0.04&a&&
2014-04-26& 56743& 41   &  1.20&  0.07&a\\
2014-04-26& 56743& 41   &  0.91& 0.30&e&&
{\bf 2014-04-17}& {\bf 56764}& \bf 9.0& \bf   3.50&  {\bf 0.08}&\bf b\\
2014-05-10& 56787& 41   &  1.15 & 0.05&e&&
\bf 2014-05-17&\bf 56794&\bf 5.5&\bf    4.50&\bf 0.24&\bf b\\
\bf 2014-05-26&\bf 56803&\bf 5.5&\bf    3.90&\bf 0.09&\bf b&&
2014-05-31& 56808& 21.2 &  4.21&  0.17&a\\
2014-05-31& 56808& 32   &  2.90&  0.13&a&&
2014-05-31& 56808& 41   &  2.94&  0.12&a\\
2014-05-31& 56808& 41   &  3.5 & 0.4 &e&&
2014-08-23& 56892& 15 &    0.63 & 0.013& g\\
2014-08-30& 56899& 23 &    0.26 & 0.006&g&&
2014-08-30& 56899& 43 &    0.15 & 0.003&g\\
\bf 2015-02-20&\bf 57073&\bf 9.0&\bf 3.00&\bf 0.30&\bf b&&
\bf 2015-09-11&\bf 57276&\bf 33.0&\bf   1.80\bf&  0.05&\bf b\\
\bf 2015-09-16&\bf 57281&\bf 44.6&\bf   2.45&\bf  0.05&\bf b&&
\bf 2016-04-23&\bf 57502&\bf 343&\bf    2.80&\bf  0.23&{\bf b}, d\\
2016-07-12& 57581& 44.2&   5.79&  0.05&f&&
2016-07-15& 57584& 226&    4.70&  1.29&f\\
\bf 2016-08-30&\bf 57630&\bf 343&\bf    3.10&\bf   0.11 &{\bf b}, d&&
\bf 2016-08-31&\bf 57631&\bf 343&\bf    3.29&\bf   0.12 &{\bf b}, d\\
\bf 2016-09-08&\bf 57639&\bf 343&\bf    3.90&\bf   0.13 &{\bf b}, d&&
\bf 2017-10-06&\bf 58032&\bf 225&\bf    1.41&\bf   0.08&{\bf b}, c\\
\bf 2017-10-07&\bf 58033&\bf 225&\bf    1.53&\bf   0.08&{\bf b}, c&&
\bf 2017-10-09&\bf 58035&\bf 225&\bf    0.79&\bf   0.07&{\bf b}, c\\
\bf 2017-10-10&\bf 58036&\bf 225&\bf    1.22&\bf   0.06&{\bf b}, c&&
\bf 2017-10-11&\bf 58037&\bf 225&\bf    1.25&\bf   0.07&{\bf b}, c\\
\bf 2017-10-12&\bf 58038&\bf 225&\bf    1.44&\bf   0.07&{\bf b}, c&&
\bf 2017-10-14&\bf 58040&\bf 225&\bf    3.16&\bf   0.08&{\bf b}, c\\
\bf 2017-10-17&\bf 58043&\bf 225&\bf    5.34&\bf   0.09&{\bf b}, c&&
\bf 2017-10-18&\bf 58044&\bf 225&\bf    5.42&\bf   0.11&{\bf b}, c\\
\bf 2017-10-20&\bf 58046&\bf 225&\bf    5.60&\bf   0.12&{\bf b}, c&&
\bf 2019-09-08&\bf 58734&\bf 5.5&\bf    $<0.50^\dagger$&\bf\dots&{\bf b}\\
\bf 2019-09-21&\bf 58747&\bf 9.0&\bf    0.40&\bf  0.10&{\bf b}&&
\bf 2019-09-20&\bf 58748&\bf 318&\bf    1.32&\bf  0.15&{\bf b}\\
\bf 2020-11-27&\bf 59180&\bf 9.0&\bf    0.36&\bf  0.06&{\bf b}&&\\
\hline\\
\end{tabular}
\begin{tabular}{p{0.95\textwidth}}
{\footnotesize 
$^\dagger$A 3-$\sigma$ upper limit.
Reference:
a. \cite{yus2015}.
b. This paper.
c. The ALMA observations at the ten epochs were carried out by the PI Tsuboi in \citep{tsu2017}. 
   We re-processed the ALMA archival data (2017.1.00500.S) and determined the flux densities of 
   SGR J1745-2900 at 225.75 GHz. 
d. Detected in the ALMA image made from observations on 2016-4-23, 2016-8-30/31, and 2016-9-8 \citep{tsu2017}.
   We re-processed the ALMA archival data (2015.1.01080.S) and determined the flux densities of 
   SGR J1745-2900 from each of the 340-GHz images at the four epochs.
e. Mean values determined from two VLA Q-band measurements with a baseline filter ($>500$ k$\lambda$) 
   and all data \citep{gel2017}; for the case of std = 0, the smaller error in the two measurements 
   is adopted.
f. \cite{yus2017}. 
g. \cite{bow2015}.
}
\end{tabular}
\end{table*}

\begin{figure}[!h]
\centering
\includegraphics[angle=0,width=100mm]{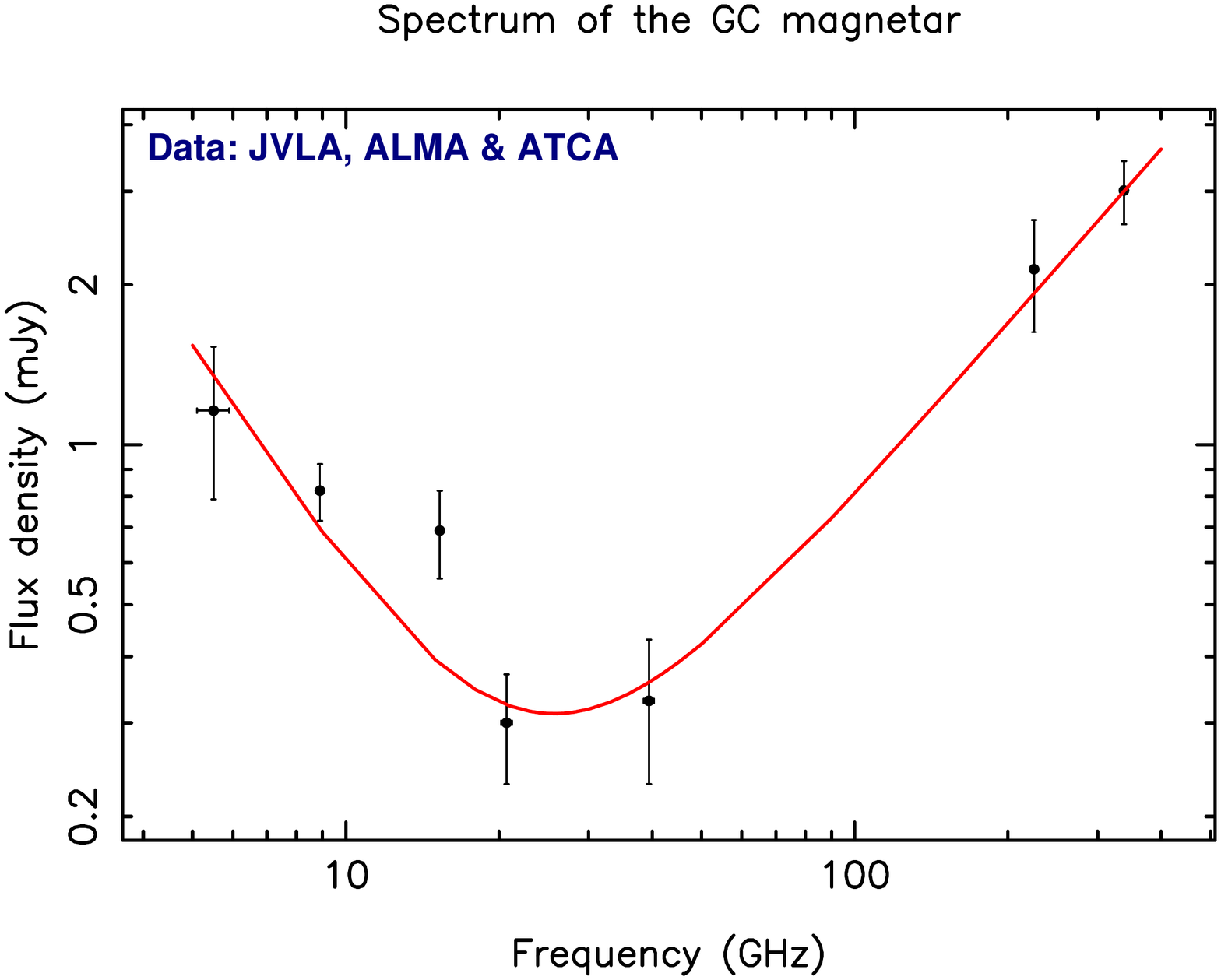}
\caption{
Averaged spectrum of SGR J1745-2900, made with band-averaged flux densities observed
with the JVLA and ALMA (values summarized in Table 3) as well as with ATCA \citep{sha2013}.}
\label{fig6}
\end{figure}

\subsubsection{Radio spectrum of the GC magnetar}
In addition to the JVLA and ALMA data, we also collected the flux-density data of the magnetar 
from ATCA \citep{sha2013}, yielding a total of 91 flux-density measurements of this object at 
radio wavelengths over the 6.5 yrs since the 2013 outburst. To investigate the spectrum  of the 
magnetar, we binned the data into nine bands with frequency ranges corresponding to those 
used for the ALMA and JVLA observing bands. We then computed the mean flux density 
($\overline{\rm S}_\nu$) and variance ($\sigma_\nu^2$) in each of the bands, weighted 
by $wt(i) = \sigma(i)^{-2}$ where $\sigma(i)$ is the uncertainty of each individual 
flux density, $i$. For non-detections, a zero weight was adopted, assuming 
that the actual uncertainties of the non-detections due to the errors 
in the calibration for the system and
atmostheric issues are much greater 
than the cited rms errors. For a total of $n$ measurements in each bin, the number of
non-zero weighted data points,  $m$, is less than or equal to $n$, (${ m\le n}$), 
within each bin. The error of the mean flux density can be determined with 

$$\displaystyle \sigma_{\overline {\rm S}_\nu}=\sqrt{\sigma_\nu^{2}\over { m}}.$$    
\parindent 0pt

We note that the Ka band shows a lower detection rate ($\sim$55\%) than the Q-band ($\sim$68\%),
indicating that a spectral minimum is located near the Ka-band frequency of 33 GHz, given the 
fact that the rms noise at the Ka-band center frequency of 33 GHz   
is about a factor of 2 smaller than that of Q-band (see Table 1).
To reduce the bias due to zero-weighting on the data with negative detections, we combined 
the Ka-band and Q-band bins' data and re-computed the mean flux density and its uncertainty. 
Table 4 summarizes the results. Fig. 6 shows the averaged spectra  observed with the JVLA, ALMA 
and ATCA over 6.5 yrs since the outburst. The vertical bars on each point reflect the large 
intrinsic variation of the flux density from SGR J1745-2900 over the observed time scale.
However, a spectral minimum appears to be present at a frequency around 30 GHz, hereafter 
referred to as the transition frequency $\nu_{\rm t}$, which appears to separate the spectrum 
into two components arising from two emission regimes at centimeter and millimeter-submillimeter 
wavelengths. The flux density of the centimeter-wave component is typically $\sim$1 mJy 
while the millimeter-wave component is about three times more intense.

\begin{table}[!h]
\footnotesize
\centering
\tablenum{4}
\setlength{\tabcolsep}{1.7mm}
\caption{Mean flux densities of SGR J1745-2900 in the JVLA/ALMA bands }
\begin{tabular}{lrrrc}
\hline\hline \\
{Band code} &
{$n$}       &
{$m$}       &
{$\overline\nu \pm \Delta\nu$ (GHz)}&
{$\overline {\rm S}_\nu \pm \sigma_{\overline{\rm S}_\nu}$ (mJy)}
\\
(1)&(2)&(3)&(4)&(5)\\
\hline \\
A7& 5& 5&338.0$\pm$5.0&3.01$\pm$0.41\\
A6&11&11&225.1$\pm$ 0.1&2.14$\pm$0.51\\
Q-Ka&34&23&39.4$\pm$ 0.9&0.33$\pm$0.10\\
K &15&10& 20.7$\pm$ 0.5&0.30$\pm$0.07\\
Ku& 7& 7& 15.3$\pm$ 0.2&0.69$\pm$0.13\\
X &12&12&  8.9$\pm$ 0.1&0.82$\pm$0.10\\
C &7& 6&  5.5$\pm$ 0.4&1.16$\pm$0.37\\
\hline\\
\end{tabular}
\begin{tabular} {p{0.40\textwidth}}
{\scriptsize
The band-averaged data are derived from the individual flux-density measurements with
the JVLA and ALMA (Table 3) as well as ATCA \citep{sha2013}.}
\end{tabular}
\end{table}

\parindent=3.5mm 
We then carried out a least-squares  fitting  of the spectrum to both the cm- and 
mm-components with two power-law functions. The centimeter data are well fit with 
a power-law spectral index of $\alpha_{\rm cm} = -1.5\pm0.6$ 
($S_{\nu}\propto\nu^{\alpha_{\rm cm}}$), a steep spectrum similar to that of radio 
pulsars. A steep spectrum of $\alpha=-1$ was previously reported between 4.5 and 
8.5 GHz for the observations during the 2013-5-30 flare of the cm-component \citep{sha2013}. 
The spectral data points rising through millimeter wavelengths can be fit with a 
spectral index of  $\alpha_{\rm mm} = 1.1\pm0.2$, indicating the presence of 
an  emission bump or a plateau at higher frequencies. Such a high-frequency 
plateau appears not to be unique to the GC magnetar. The radio-active magnetar 
1E 1547.0$-$5408 has also been observed to have a  spectrum rising towards millimeter
wavelengths \citep{chu2021}. Overall, the  spectrum of the GC magnetar at the 
frequencies in the range between 5 and 310 GHz is described with a combination 
of the two power-law functions (see Fig. 6) with a transition frequency 
$\nu_{\rm t} \approx 30$ GHz corresponding to a minimum flux density of 
$\sim$0.3 mJy.

\begin{figure}[!h]
\centering
\includegraphics[angle=0,width=85mm]{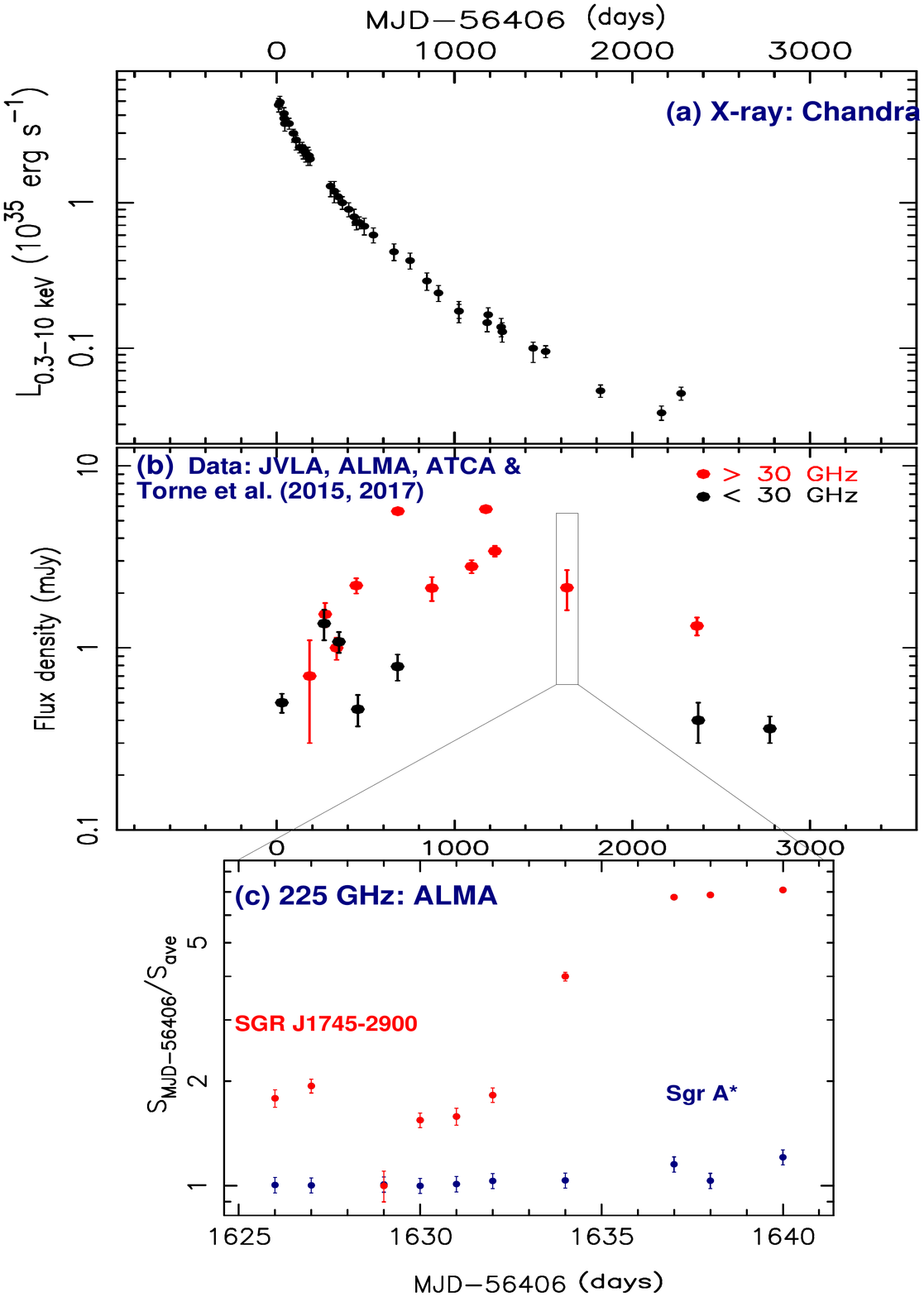}
\caption{ (a) The X-ray luminosity variation after the onset of outburst of SGR J1745-2900,
observations by the Chandra X-ray observatory \citep{rea2020}. (b) 100-MJD bin averaged
radio light curve of the magnetar J1745-2900 for both the mm- (red dots) and  cm- (black dots)
components. In addition to the flux-density measurements from the JVLA and ALMA summarized
in Table 3, other telescope data from literatures \citep{sha2013,tor2015,tor2017} are
included. (c) Detail of ALMA observations in the interval between 2017-10-6 and 2017-10-20
(1626 days after the outburst of SGR 1745-2900 on 2013-4-23) show a day-to-day variability
of the magnetar SGR J1745-2900 at 225 GHz  displayed as normalized flux
densities S$_{\rm MJD-56406}$/S$_{min}$ (red dots), where the minimum flux density
S$_{min}=$0.79$\pm$0.08 mJy for the magnetar on MJD 58035. The variability of SGR J1745-2900
during the two-week interval is compared with that of Sgr A* (dark-blue dots); Sgr A*
(S$_{min}$=3.015$\pm$0.013 Jy and  S$_{ave}$= 3.26$\pm$0.09 Jy) that is three orders in
magnitude more luminous than the magnetar.
}
\label{fig7}
\end{figure}

\subsubsection{Radio variability of SGR J1745-2900}
We inspected the radio variability of the GC magnetar SGR J1745-2900, including    
a total of 161 flux-density measurements since the onset of the 2013 outburst.
The measurements by \cite{tor2015,tor2017} are also included in addition to 
the JVLA-ALMA (Table 3) and ATCA \citep{sha2013} data. The magnetar appears 
to be highly variable on all observed time scales and wavelengths 
\citep[e.g.][and this paper]{sha2013, yus2015}. We utilized a bin-averaging 
algorithm similar to the analysis of the magnetar spectrum (Sec.4.3.2).
We binned the MJD or the time axis with a constant interval of 100 days. 
The two spectral components $--$ low-frequency and high-frequency $--$
corresponding to the cm-component ($\nu<30$ GHz) and the mm-component ($\nu>30$ GHz), 
were examined separately. We then computed the weighted mean flux densities and 
the corresponding dispersions in each of the 56 MJD-bins for both mm- and cm-components. 
The results for the non-empty bins are tabulated in Table 5. Columns 1 and 2    
show the bin ID (binID) and the corresponding central MJD. Columns 3 and 4 are  
the numbers of all the measurements ($n$) and the non-zero weight data ($m$) 
in the corresponding MJD-bins; as above, the measurements with only upper limits 
are zero-weighted. Column 5 is the mean MJD of the observing dates. Column 6    
gives the  mean flux-density ($\overline{S}_{\rm binID}$), weighted by the 
inverse variance on each of the measurements, along with the uncertainty of
the mean $\sigma_{\overline{S}_{\rm binID}} = \sqrt{\sigma^2_{\rm binID}/{ m}}$.
And the standard deviation ($\sigma_{\rm binID}$), or the dispersion due mainly 
to the variation in flux density, is listed in column 7.

Fig. 7b  plots the MJD-bin averaged radio light curves for the two spectral components, 
red for mm  and black for cm. Unlike the X-ray light curve (Fig. 7a)  observed by the 
Chandra X-ray observatory \citep{rea2020} that shows a smooth decrease in X-ray luminosity, 
both the cm- and mm-components varied significantly in the first 800 and 1700 days. Our 
two measurements of flux density based on the earlier JVLA observations  at 5.5 GHz
showed a significant variation from 4.50$\pm$0.24 mJy on 2014-05-17 to 3.90$\pm$0.09 mJy 
on 2014-05-26, on the time scale of a week, which is consistent with the large variability at 5 GHz 
observed during May 2013 \citep{sha2013}.

The two most recent JVLA measurements at 9 GHz on 2020-11-27 (MJD 59180) and
2019-09-21 (MJD 58777)  indicate that the flux density of the cm-component  
decreased to 0.4 mJy from 3.5 mJy in the JVLA observation on 2014-04-17 (MJD 56764).
Also, we failed to detect the source at 5.5 GHz based on the JVLA observations on 
2019-09-08 (58734), imposing a 3-$\sigma$ upper limit of 0.5 mJy while the flux 
density was 4.5 mJy based on our JVLA observation at 5.5 GHz on 2014-5-17 (MJD 56794). 
The mean flux density of the cm-component of the magnetar decreased by a 
factor of 3 over a period of 6.5 yr (Fig. 7b and Table 5).

The Chandra X-ray observations of SGR J1745-2900  following the outburst show a 
smooth decrease from 4.9$\times10^{35}$ erg $^{-1}$ at the onset of the X-ray 
outburst to 0.047$\times10^{35}$ erg $^{-1}$ at the most recent epoch (Fig. 7a). 
The X-ray luminosity therefore dropped by two orders in magnitude over 6 years. The 
general trend of decreasing radio flux density is consistent with the X-ray 
light curve,  although the cm-component has a much slower decline, 
is much more variable, and shows a large range of variability on time scales
from days to years.

The mm-component stayed at a relatively low level, with bin-averaged flux densities of 
0.7$\pm$0.40 and 1.53$\pm$0.23 mJy during the second and third  MJD-bins after 
the outburst (Table 5 and Fig. 7b). During the first 300 days, the mm-component 
was difficult to detect (only 5 detections out of 11 observations). We 
note that the JVLA Q-band data observed during 2013 fall and 2014 spring 
were reduced by two independent groups \citep{yus2015,gel2017}. The mm-component 
reached a maximum of 5.63$\pm$0.19 mJy in the 7th MJD-bin (601-700 days after 
the outburst), based on 35 observations with a detection rate of 100\%. 

The 17th MJD bin of the mm-component (between 1600-1700 days) contains 10 individual 
ALMA observations at 225 GHz, all carried out within a two-week period during 
October 2017 with a high-resolution (0.02") configuration \citep{tsu2019}. The mean flux 
density for the mm-component in this bin was  2.25$\pm$0.56 mJy. With this 
configuration, the extended HII emission, as well as dust emission from the local 
medium, are well resolved out and only hyper-compact sources can be detected. The 
typical rms noise of the ALMA images is about 20-30 $\mu$Jy beam$^{-1}$, and the 
magnetar has S/N ratios of 50-100. At a distance of $\sim$3" from the magnetar, Sgr A*,
with a flux density near 3 Jy, is an excellent reference source to examine the 
day-to-day variability of the magnetar. The ALMA data show the magnetar to be highly 
variable at 225 GHz on a time scale of days. The source started at 1.41$\pm$0.08 mJy 
on 2017-10-6, dropped to a minimum of 0.79$\pm$0.08 mJy three days later on  
2017-10-9, and then increased by a factor 7 within 8 days reaching 5.34$\pm$0.09 mJy
on 2017-10-17, staying at that level for the next several days. To compare the 
variability of SGR J1745-2900 with Sgr A*, we normalized the source flux densities 
by their own minimum flux densities during the observing period. Fig. 7c shows 
the relative variability for both the magnetar and Sgr A* during the two weeks 
in bin 17. We define a relative variability parameter, $\mathscr{RV}$, to 
quantitatively describe the magnitude of variability relative to a minimum flux 
density $S_{\rm min}$, given a maximum flux density $S_{\rm max}$ of a target 
source observed in a period:

$$ \mathscr{RV} ={\left[{\rm S_{\rm max}- S_{\rm min}}\right]\over{\rm S_{\rm min}}}.$$

\parindent 0pt
The magnetar shows the relative variability $\mathscr{RV}\approx$  6 while 
Sgr A* is observed to be only moderately variable in the same ALMA program, 
with $\mathscr{RV}\approx$  0.2..  

\parindent=3.5mm
Finally, the latest available detections of the magnetar in 2019-2020, are 1.32$\pm$0.15 mJy 
at 318 GHz with ALMA on 2019-09-20 (MJD 58748) and 0.36$\pm0.06$ mJy with the JVLA on 
2020-11-27 (MJD 59180).

\begin{table}[!htp]
\footnotesize
\centering
\tablenum{5}
\setlength{\tabcolsep}{1.7mm}
\caption{Mean flux densities of SGR J1745-2900 in 100-day MJD bins}
\begin{tabular}{lcrlccc}
\hline\hline \\
{binID}   &
{MJD}     &
{$n$}       &
{$m$}       &
{$\overline {\rm MJD}$ }&
{$\overline {\rm S}_{\rm binID} \pm \sigma_{\overline{\rm S}_{\rm binID}}$} &
{$\sigma_{\rm binID}$}\\ 
          &
(day)     &
          &
          &
(day)     &
(mJy)     &
(mJy)     \\
\\
(1)&(2)&(3)&(4)&(5)&(6)&(7)\\
\hline \\
\multicolumn{7}{c}{Centimeter-wave component} \\
 1&56456& 16& 16& 56435.1&0.50$\pm$0.06&0.24\\
 3&56656&  5&  3& 56673.2&1.36$\pm$0.26&0.45\\
 4&56756& 10& 10& 56755.7&1.08$\pm$0.14&0.44\\
 5&56856& 12& 12& 56862.4&0.46$\pm$0.09&0.31\\
 7&57056& 10& 10& 57085.9&0.79$\pm$0.13&0.41\\
24&58756&  2&  1& 58777.0&0.40$\pm$0.10&0.10\\
28&59156&  1&  1& 59180.0&0.36$\pm$0.06&0.06\\
  &     &   &   &                &             \\
\multicolumn{7}{c}{Millimeter-wave component} \\
 2&56556&  3&  1&56591.5&0.70$\pm$0.40&0.40\\
 3&56656&  9&  5&56678.6&1.53$\pm$0.23&0.51\\
 4&56756& 10& 10&56742.4&1.00$\pm$0.14&0.44\\
 5&56856& 26& 21&56853.2&2.20$\pm$0.21&0.96\\
 7&57056& 37& 37&57087.3&5.63$\pm$0.19&1.16\\
 9&57256&  2&  2&57278.5&2.13$\pm$0.32&0.45\\
11&57456&  1&  1&57502.0&2.80$\pm$0.23&0.23\\
12&57556&  2&  2&57582.5&5.78$\pm$0.04&0.06\\
13&57656&  3&  3&57633.3&3.40$\pm$0.23&0.40\\
17&58056& 10& 10&58038.4&2.14$\pm$0.53&1.68\\
24&58756&  1&  1&58770.0&1.32$\pm$0.15&0.15\\
\hline\\
\end{tabular}
\begin{tabular} {p{0.45\textwidth}}
{\scriptsize
The 100-day-averaged data are derived from a total of 161 individual flux-density 
measurements with the JVLA, ALMA (Table 3), ATCA \citep{sha2013} and other 
single-dish telescopes \citep{tor2015,tor2017} since the onset of the outburst 
at the epoch 2013-4-23 (MJD 56406).}
\end{tabular}
\end{table}

\section{Astrophysical implications}
A population of hyper-compact radio sources (HCRs) are detected at 33 and 44.6 GHz 
with the JVLA in the vicinity of Sgr A* within a radius of 13 arcsec. The new survey 
was motivated by the previous JVLA detections of 110 Galactic center compact radio 
sources (GCCRs) at 5.5 GHz in the radio bright zone within a radius of 7.5 arcmin 
from Sgr A* but outside Sgr A West.

\subsection{ Spectral types of HCRs \& their distribution in flux-density}
Fig. 8 shows the flux-density distribution of the HCRs as compared with that of the GCCRs.
The distribution of the GCCRs is similar to the  high-luminosity tail of the pulsars'
distribution in the Galactic disk \citep{kra1998,man2005,zmg2020}. The  HCRs have a
relatively narrow distribution from $-1$ to 0.6 in  Log(S [mJy]), or ranging  from 0.1
to 4 mJy, peaked at $-0.4$ in Log(S [mJy]) or 0.4 mJy. We note that the flux density of
the peak in the HCR distribution is close to the minimum value 0.32 mJy at the transition
frequency observed in the band-averaged spectrum for the GC magnetar J1745-2900
(see Sec.4.3.2).

We divided the HCRs into three sub-types according to their spectral indices
$\alpha_{44.6/33}$  ($S\propto \nu^{\alpha_{44.6/33}}$)  derived between 44.6 and 33 GHz:
flat ($0.2\ge\alpha_{44.6/33}> -0.5$), steep ($\alpha_{44.6/33} \le -0.5$) and inverted
($\alpha_{44.6/33}>0.2$).

Of all the HCRs, 58\% (38/65) are steep-spectrum sources, 26\% (17/65) are inverted-spectrum
sources, and 15\% (10/65) have a flat spectrum. We note that HCR49 is a double; so a total
of 65 HCR components are included in the spectral-index analysis. The inset of Fig. 8 
shows the flux-density distributions for each of the three spectral types with a finer bin,
$\Delta$Log(S [mJy]) = 0.2.

\subsubsection{Flat-spectrum HCRs}

The flux density distribution of the flat-spectrum HCRs appears to be 
uniform within the range from 0.16 to 1.6 mJy. Some fraction of the flat-spectrum HCRs
might be unresolved peaks in the HII region components since a flat spectrum at Ka-
and Q-bands is characteristic of free-free emission in optically thin HII regions.

\subsubsection{Steep-spectrum HCRs}
This sub-sample consists of 38 members, the largest sub-sample among the three,
in which the flux densities are statistically well distributed. The distribution
can be fitted with a Gaussian, with a mean value of $\mu=-0.35$ and a standard
deviation of $\sigma=0.22$ in Log(S [mJy]). The mean value of the steep-spectrum
HCRs corresponds to 0.45 mJy in flux density.

The spectral index $\alpha_{44.6/33}$ of the steep-spectrum HCRs
is in the range between $-6.5$ and $-0.6$, giving a mean $\mu_\alpha=-1.8\pm0.2$.
The steep spectrum of  this sub-sample differs distinctly  from the HII components,
suggesting the presence of a population of hyper-compact nonthermal radio sources in the
central parsec. The nonthermal HCRs are likely
associated with the massive stellar remnants that are expected to be distributed in
the close vicinity of Sgr A* \citep{mor1993,hai2018,gen2018}.

\subsubsection{Inverted-spectrum HCRs}

For this sub-sample, the mean flux density and standard deviation are
$\mu=0.5$ mJy and $\sigma=0.25$ mJy, respectively. The spectral index
$\alpha_{44.6/33}$ of the inverted-spectrum HCRs is in the range between 0.21 and 1.65,
with a mean value of $\mu_\alpha=0.61\pm0.08$. Among the three spectral sub-samples,
the distribution of the inverted-spectrum HCRs appears to most closely follow the
GCCR distribution, which matches the high-luminosity tail of the distribution that
normal pulsars would have at the Galactic center \citep{zmg2020}.   
Normal pulsars  usually have a steep spectrum
at centimeter wavelengths and are difficult to detect at high frequencies. However, the
GC magnetar, emitting at the Ka- and Q-bands, falls into this sub-sample. As shown
in Sec. 4.3.2, an inverted spectral component of SGR J1745-2900 is present at
millimeter wavelengths in the averaged spectrum of a large sample of observations. The
inverted spectrum cannot be simply attributed to time variability of the flux density.
By analogy with SGR J1745-2900, the inverted spectral component could be
indicative of an association of the HCRs with magnetars.

On the other hand, hyper-compact HII regions associated with a late-type massive star may
also have a similar characteristics in spectral index, flux density level and compactness.
For example, a group of compact components have been detected at mm-submm wavelengths
in the IRS 21 complex (see Fig. 2b). The hyper-compact sources associated with IR emission
more likely belong to hyper-compact HII or HC HII regions associated with young stellar objects
rather than old massive stellar remnants.  The hyper-compact radio components of IRS 21 are
therefore not included in the HCR catalog (Table 2) in this paper.
IRS 21 will be discussed  in a separate paper.

From cross-correlation between the HCRs of this paper and the nine 7mm-IR sources
that are found to be associated with strong stellar winds \citep{yus2014}, we find that six
of the nine candidate IR stars have HCR counterparts. The spectral indices of the six
HCRs -- five with inverted spectra and one with a flat spectrum -- are consistent with the radio
emission produced by the ionized winds of hot, massive stars \citep{pan1975}. Thus, a
large fraction of the 17 inverted-spectrum HCRs might consist of thermal free-free
emission sources. Apparently, a significant portion of the inverted-spectrum
HCRs is associated with late-type massive stars.

We speculate that more magnetars besides SGR J1745-2900
reside in our inverted-spectrum sample of HCRs. However, 
we must be able to distinguish
such objects from  compact radio sources associated with young massive stars.
Further high-resolution ALMA observations of the variability and polarization
characteristics of this subsample will be crucial for identifying the nonthermal nature of
candidate magnetars and stellar-mass black holes.

\begin{figure}[!h]
\centering
\includegraphics[angle=0,width=95mm]{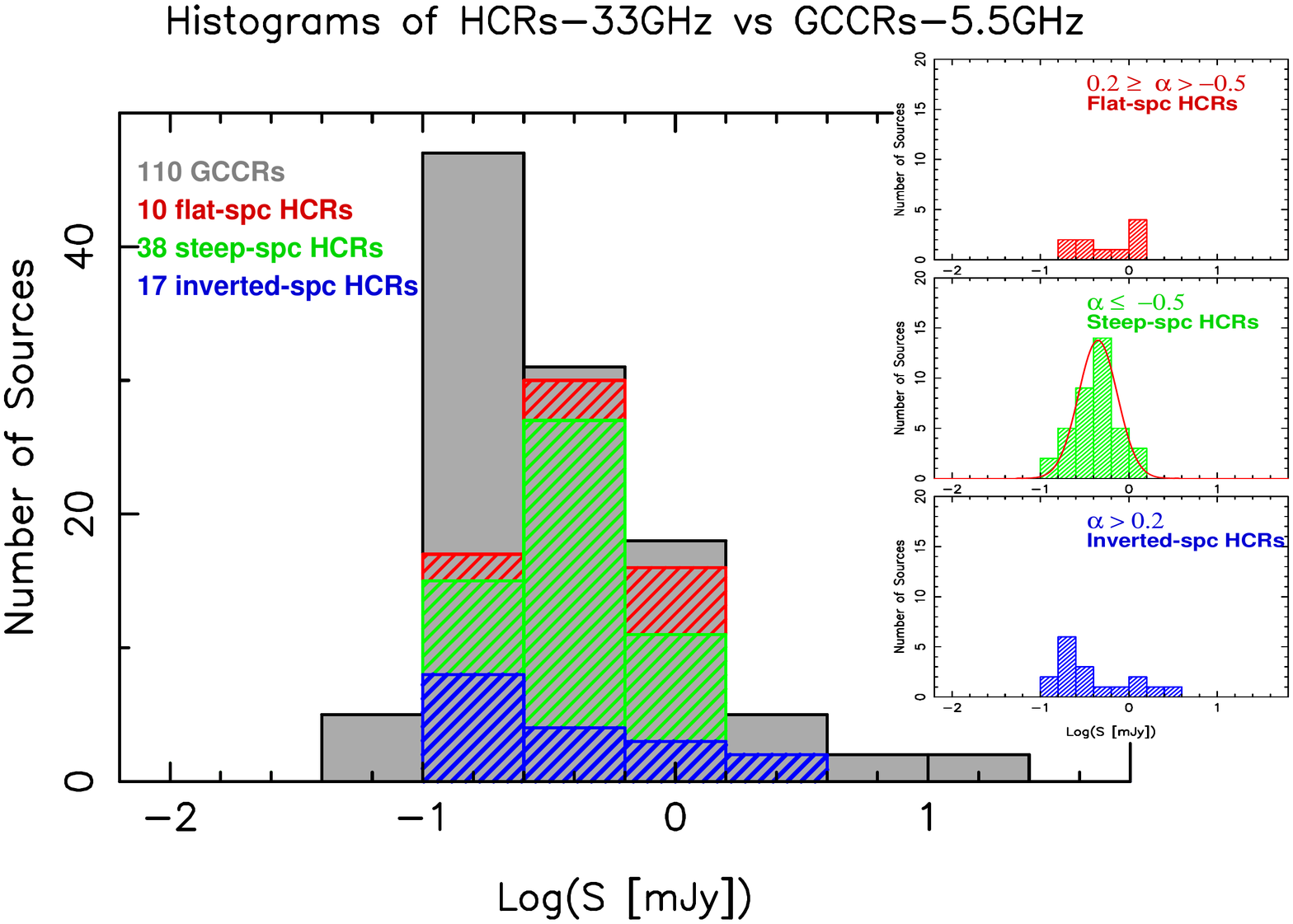}
\caption{Flux-density distribution of the HCRs at 33 GHz within the central
13"$\times$13" region versus the GCCRs \citep{zmg2020} detected in the GC radio
bright zone (15'$\times$15'). The grey histogram represents the 110 GCCRs detected
at 5.5 GHz outside Sgr A West. The 64 HCRs (65 components, HCR49 is double), shown
by the hatched histograms, are divided into three types: flat $-0.5<\alpha\le0.2$
(red), steep $\alpha\le-0.5$ (green) and inverted $\alpha\ge0.2$ (blue), according
to their radio spectral index determined between 44.6 and 33 GHz, except for three
steep-spectrum sources -- HCR22, HCR32 and HCR64 -- which were not detected in the Q-band
data used in this paper. The spectral indices for HCR22 and HCR32 are determined
using X-band data discussed in this paper. The spectral index for HCR64, the
micro-quasar, is determined using the K-band data of \cite{zmga2009} assuming that
the source was in a quiescent state during the 2015 observations. The inset panels
on the right are the more finely binned individual distributions of the three 
spectral sub-samples. The distribution of the steep-spectrum HCRs is fitted 
with a Gaussian (red curve).
}
\label{fig8}
\end{figure}

\begin{figure}[htp]\includegraphics[angle=0,width=100mm]{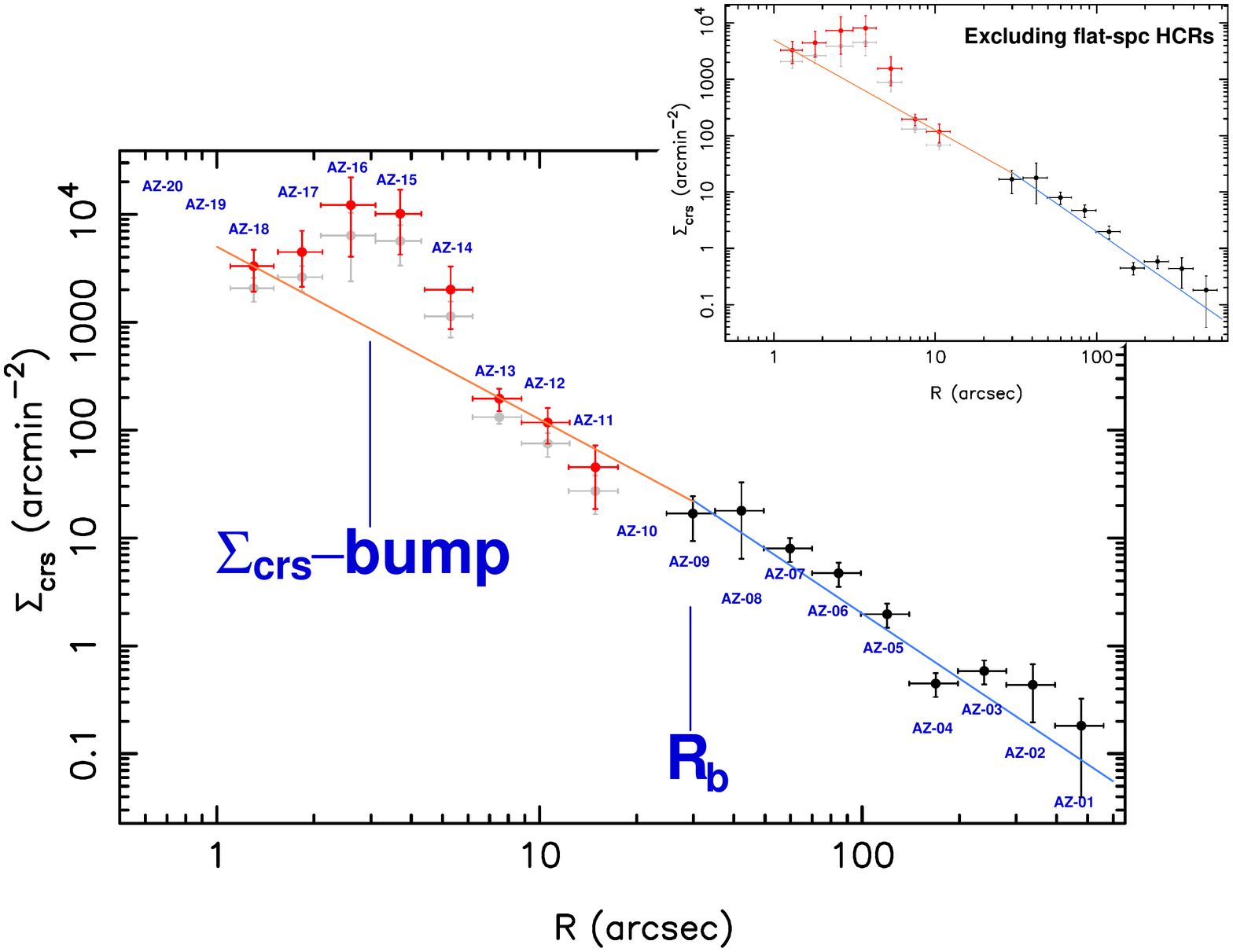}
\caption{Radial distribution of the surface density of compact radio
compact sources detected at the Galactic center. The red points are
the surface density data $\Sigma^*_{\rm crs}$
corrected for a higher equivalent sensitivity cutoff at 5.5 GHz by multiplying
the original 33-GHz surface density data, $\Sigma_{\rm crs}$ (light-grey points),
by a correction factor, $\eta =2.4$, derived from the 33-GHz observations of the
HCRs (see Appendix A of this paper).  The black points mark the data of GCCRs
observed at 5.5 GHz. In the annular zones around $R=3"$, a bump in $\Sigma_{\rm crs}$
is present.  The orange line is the best fit using a power-law with an 
index $\Gamma=1.6\pm0.2$
for the inner region outside of the bump at 1.5 - 7 arcsec; and
the blue line indicates the least squares fitting to the steeper power law with
$\beta=2.0\pm0.2$ for the outer region. The break radius, $R_b$, is
$\sim30"$, or $\sim1.2$ pc.  The top-right inset shows the surface-density
distribution of the HCRs excluding the flat-spectrum sources.
}
\label{fig9}
\end{figure}

\subsection{A dense group of HCRs and the radial distribution of their surface-density}

The surface density of 110 GCCRs, located outside Sgr A West ($> 1$ arcmin from Sgr A*)
but within the RBZ ($< 7.5$ arcmin from Sgr A*), is $\sim$0.6 counts arcmin$^{-2}$.
This surface density is an order of magnitude greater than that of background
extragalactic compact radio sources \citep[e.g.][]{con2012,gim2019}. The dense
group of 64 HCRs, located within a radius of 13" from SgrA*, has a relatively higher
surface density, with an average value of $\sim$500 counts arcmin$^{-2}$.
We conclude that contamination of our sample by extragalactic background sources
is very likely to be negligible.    

In Appendix A, we developed a procedure to construct the surface-density ($\Sigma_{\rm crs}$) 
distribution of the radio compact sources (RCS) as function of the projected 
radial distance ($R$) from Sgr A*, including 64 HCRs at $R<13$" (this paper) 
and the 110 GCCRs outside Sgr A West \citep{zmg2020} as well as the Galactic 
center transient \citep[GCT, see][]{zha1992}. Fig. 9 shows the radial distribution
of the surface-density of the RCSs detected within the RBZ. Excluding the four 
data points related to a $\Sigma_{\rm crs}-$bump 
in the radial distance range between 1.5" and 7", the surface-density distribution 
can be fitted with two power laws (see Fig. 8)  as described by the ``Nuker'' 
law \citep{lau1995,gen2003,fri2016,sch2018}. In the inner region at $R <30$" but excluding
the range between 1.5" and 7",
the surface-density  distribution of the HCRs follows a power law:

\begin{equation}
\centering 
\Sigma_{\rm HCR}(R<30") \approx (5.0\pm1.8)\times10^3R_{\rm arcsec}^{-\Gamma}~~~~ ({\rm arcmin^{-2}})
\end{equation}
where the index $\Gamma=1.6\pm0.2$. At  large radii $R\ge30$", the surface-density 
distribution of the GCCRs shows a steeper power law:

\begin{equation}
\centering
\Sigma_{\rm GCCR}(R\ge30") \approx (2.0\pm0.9)\times10^4R_{\rm arcsec}^{-\beta}~~~~ ({\rm arcmin^{-2}})
\end{equation}   
where $\beta=2.0\pm0.1$. At $R\sim30$", the two power laws intersect.
\vskip 5pt
We note that the power-law distribution derived from HCRs is much steeper than that $\Gamma\sim0.6$
of the stellar cusp determined with the $K_{\rm s}$ stars in magnitude range between
12.5 and 18.5 in the radial distance range between 1" and 50" \citep{gal2018}.

\subsection{Candidate massive stellar remnants}

\subsubsection{Mass segregation}

It has been known for several decades that the central parsec contains a large number of 
young massive stars \citep{kra1991,kra1995,pau2001,pau2006,gil2009,lu2009,bar2010,lu2013}, 
with $\sim$100 O and Wolf-Rayet (WR) stars confined within a radius of $\sim$0.4 pc 
\citep{gen2003,ghe2005}. These stars are relatively young ($\le10$ Myr) and are orbiting 
Sgr A*, the supermassive black hole (SMBH) at the Galactic center. The early-type stars  
provide substantial UV photons to maintain the ionization of the gas within the central 
parsec \citep{zha2010}.

The stars in the central cusp apparently have a top-heavy present-day mass
function \citep[over-abundance of high-mass stars, see][]{bar2010,lu2013}; and the fact
that the orbits of many of them collectively define a coherent disk suggests that the young
nuclear cluster may have originated via {\it in situ} star formation \citep{lev2003,lev2007,nay2007,col2008}.
The process that formed the young nuclear cluster near the central black hole is likely to
have been a recurring one. This process follows a limit cycle of activity wherein the star
formation is a violent event coinciding with a heavy accretion episode
onto the SMBH. Thus the inner disk is quickly disrupted and dissipated in the process.
Following the disruption of the inner disk, continued migration of gas from the central molecular zone toward
the center rebuilds the disk, eventually leading to the next star formation event \citep{mgb1999}.
With repeated instances of this cycle, the remnants of massive stars are produced
and left in place to collect at the bottom of the Galaxy's gravitational potential well.
Furthermore, the more massive remnants, particularly stellar-mass black holes,
will undergo mass segregation in the dense central star cluster, as a result of dynamical
friction, and become even more concentrated toward the black hole
\citep{bah1976,mor1993,mir2000,pfa2004,fak2006,ale2009,mer2010,ant2012,ale2017,gen2018}.
The mass segregation may partially explain the intense flux of gamma rays and X-rays from the
Galactic center caused by the possible presence of a large population of millisecond pulsars
(MSPs) and cataclysmic variables (CVs), as suggested by \citep{arc2018}. The excess of
GeV gamma rays toward the Galactic center may alternatively be explained by a high
supernovae rate, leading to the production of neutron stars and ultimately to a MSP population
\citep{ole2015,ole2016,cal2016}.

In the case of  stellar-mass BHs that are significantly more massive
than the mean stellar mass expected for an evolved population, such heavier objects
would migrate toward the center and be distributed in a compact cluster around
the SMBH. The nature of the mass segregation depends on the relaxational coupling parameter
$\Delta =4 n_{\rm BH} M^2_{\rm BH}/\left[n_{\rm st} M^2_{\rm st} (3 + M_{\rm BH}/M_{\rm st})\right]$
for  black hole (BH) mass  $M_{\rm BH}$ and spatial number density $n_{\rm BH}$
along with stellar (ST) mass $M_{\rm st}$ and spatial number density $n_{\rm st}$.
The value of $\Delta$ is a measure of the importance of BH-BH scattering relative to
BH-ST scattering for the dynamical friction process \citep{ale2009}. Weak segregation
occurs when $\Delta\gg1$; and when $\Delta\ll1$, the strong segregation applies.
The latter case leads to a steeper slope in the 3D radial density distributions of
BHs and a larger central concentration of BHs relative to that of stars: for power-law
indices\footnote{Throughout this paper, the lower-case $\gamma$ is used for power-law
indices of 3D radial density distributions while the upper-case $\Gamma$ stands for
power-law indices of 2D radial density, or the surface density, distributions.}
$\gamma_{\rm BH}$ and $\gamma_{\rm ST}$ for BHs and stars, respectively,
$2\lesssim\gamma_{\rm BH} \lesssim 11/4$ and
$3/2\lesssim\gamma_{\rm ST} \lesssim 7/4$
($\gamma_{\rm BH}-\gamma_{\rm ST}\backsimeq1$) \citep{ale2017}.
Long-lived stellar populations usually have $\Delta<0.1$, and the Galactic center
is expected to be strongly segregated \citep{mor1993}.
\cite{ale2009} have shown
the effects of strong mass segregation on the density distribution of a model
stellar population around Sgr A* (mass $M_{\rm SMBH} = 4\times10^6$ M$_\odot$);
the modelled population includes main sequence stars (MS$\sim1$ M$_\odot$) and
stellar remnants, including white dwarfs (WD$\sim0.6$ M$_\odot$), neutron stars
(NS$\sim1.4$ M$_\odot$) and black holes (BH$\sim 10$ M$_\odot$). Their study
demonstrates that the heavier objects produce steeper density distributions
via the mass segregation process.

We note that, in the radial range  0.04-1 pc, the power-law index
of the 2D surface-density profile of the HCRs is $\Gamma_{\rm HCR}=1.6\pm0.2$
(this paper) which is much greater than the corresponding value of $\Gamma_{K_{\rm s}}$  
for the $K_{\rm s}$ stars in  the same radial range  \citep{gal2018}.
The difference between  the power-law indices ($\Gamma_{\rm HCR}-\Gamma_{K_{\rm s}}$)
is $\sim$1.
The flat power-law profile for $K_{\rm s}$ stars provides evidence for the presence
of a dynamically relaxed stellar cusp at the Galactic center \citep{gal2018,sch2018}.
On the other hand, with the arguments presented in the previous paragraph,
the dense group of HCRs reported in this paper, with its much steeper radial
distribution (see Fig. A2) than that of the faint $K_{\rm s}$ stars, could represent
a population of massive stellar remnants that are mass-segregated in the nuclear
star cluster at the Galactic center. The HCRs are associated with active massive
stellar remnants having relatively higher radio luminosities. The surface density
profile of the HCRs (see Fig. A2) may be subject to change when a deeper radio
survey is carried out with a much lower flux density cutoff. Nevertheless, the
results from our study of HCRs in the central parsec provides evidence consistent
with the presence of a distribution of massive stellar remnants that is a steep
function of the radial distance from Sgr A*.

In addition, if the massive stellar remnants associated with the HCRs
had migrated inward via dynamical friction, the bump in surface density, 
$\Sigma_{\rm HCR}$, at 0.1-0.3 pc could be attributed to an accumulation 
in that radial range since the dynamical friction force acting on a massive 
object ceases  at roughly  half the radius of the stellar core ($\sim0.25$ pc)  
\citep{mer2010}. A maximum in the density profile  of massive remnants is 
also predicted to occure at $\sim0.2$ pc at a time $>$1 Gyr in a dynamical 
evolution model, and the bump slowly grows  and migrates inward due to the 
friction produced by fast-moving stars inside these radii \citep{ant2012}.      
The $\Sigma_{\rm HCR}$-bump may serve as an additional observational signature  
of massive stellar remnants as a consequence of stellar dynamical processes 
in galactic nuclei.

\subsubsection{Radiation spectrum}

\cite{hai2018} recently reported the identifications of a dozen low-mass back hole X-ray
binary candidates within the central parsec, implying the presence of a large population
of X-ray binaries and isolated black holes residing within that volume \citep{gen2018}.
Most of them are in a quiescent state. X-ray and radio flares and outbursts from these
stellar remnants have been discovered over the past three decades, such as the magnetar
SGR J1745-2900 \citep{ken2013,rea2020, eat2013a,sha2013,tor2015,tor2017}, the microquasar
of the X-ray transient XJ174540.0290031 \citep{mun2005, bow2005, zmga2009} and the Galactic
center transient \citep[GCT, see][]{zha1992} as well as  the X-ray PWN candidate G359.95-0.04
\citep{bag2003,wan2006,mun2008} which is likely powered by a neutron star.

With a flux-density range bewteen $\sim$0.1 and a few mJy at 33 and 44.6 GHz, 
the HCRs appear to be candidate radio counterparts of the old massive stellar 
remnants produced in the end of stellar evolution as expected. Most of them 
are in a quiescent state. Although their progenitors and ages are unknown, 
the 38 steep spectrum HCRs  determined from the JVLA observation at Ka and Q 
bands provide important clues on the nature of the radio radiation, affirming 
nonthermal radiation with a steep power law in the distribution of relativistic 
electrons. The nonthermal emission could be produced by synchrotron jets or 
outflows that were launched from the compact stellar remnants powered by 
accretion from the dense, local medium.

As described above,
one of the 17 inverted-spectrum HCRs is the GC magnetar, SGR J1745-2900, which
shows high-frequency pulses up to 291 GHz \citep{tor2017}. The continuum emission
from this magnetar has been firmly detected at high radio frequencies up to 340 GHz
\citep[][and this paper]{tsu2017}. The inverted-spectrum of the mm-component
of a magnetar towards the submillimeter appears to be a remarkable radio wavelength
signature, as predicted by the dynamical model of \citet{bel2013} using a persistent flow of
electron-position plasma. The configuration of the magnetosphere
of magnetars, created by enforced electric current and radiative drag together,
is subject to two-stream instability. Consequently
a relatively hard radio spectrum that is predicted to emerge, perhaps
extending to IR/optical/UV wavelengths, and is expected to be generated because
the instability leads to a large plasma density, and thus a large plasma
frequency \citep{bel2013}. The theory also predicts a large electric current
associated with the radio-submillimeter emission from magnetars \citep{bel2013},
producing a bright radiation beam much broader than the typical pulse width of
normal pulsars with similar periods \citep{cam2006,cam2007}. A valuable next
step will be to use this theory to formulate predicted shapes of radio
spectra and pulse profiles of magnetars for comparison with observations.

The suggestion has been made that pulsars formed at or near the Galactic center
might mostly to be magnetars, given the rather highly magnetized interstellar medium of the GC
region that could produce highly magnetized massive stars. Subsequently, strongly magnetized
neutron star remnants form because of the collapse of stellar core and concentration of the 
flux-frozen magnetic field  \citep{dex2014,mor2014}.  However, because the magnetic
flux within the core of a massive star as well as within a neutron star can undergo considerable
evolution owing, for example, to dynamo action occurring inside the star and the neutron star,
\citep[e.g.,][]{dun1992,tho1993}, this scenario remains rather speculative.  In any case,
the discovery of SGR J1745-2900 makes the central parsec a potentially interesting region
to search for new magnetars. Although the recent formation of a large number of massive
stars there could have led to a population of neutron star remnants, if those remnants are
predominantly highly magnetized, then, as with magnetars in general, they would have short
lifetimes as pulsars ($\sim$ 10$^3-10^5$ yrs) because of the powerful spindown torque
associated with the interaction of the neutron star's magnetic field with the plasma in its
environment  \citep{har1999,esp2011,dex2014,kas2017}. The time period during which a
magnetar is an observable pulsar is therefore more than two orders of magnitude shorter
than the lifetime of the massive stars observed in the young nuclear cluster ($\sim$2 - 10 Myrs).
So we might therefore expect only a few (or zero) magnetars to be found as pulsars at any one
time in the central parsec if the above speculation that massive GC stars produce highly
magnetized remnants is correct. Indeed, in addition to the scatter-broadening that occurs
primarily at longer radio wavelengths, the short lifetime of strongly magnetized pulsars could
be the main explanation for the rarity of pulsars at the Galactic center. The remaining
open question is whether magnetars remain detectable as point radio continuum sources
even after they have spun down to the point at which they can no longer be detectable
as pulsars.  If so, then we might consider that some of the HCRs are in that category.

The inverted spectrum towards short wavelengths found in
1E1547.0-5408 \citep{chu2021} in addition to SGR J1745-2900 is also consistent
with the prediction of spectral hardening at short radio wavelengths
from the two-stream instability model \citep{bel2013}. The combination of
the inverted spectrum, high variability  and high polarization --
nearly 100\% for the degree of linear polarization and 15\% for
circular polarization \citep{eat2013a} -- appears to be unique to magnetars.

Of course, both the inverted and flat spectra of HCRs can also
be interpreted as self-absorbed synchrotron emission from X-ray binaries
in the hard state when the radiation is dominated by the emission from
the corona of the compact object  \citep{cor2011}.

\section{Conclusion}

Following our 5.5-GHz JVLA survey of the Galactic center compact radio sources (GCCRs)
within the radio bright zone, we have continued to explore Sgr A West using
Ka- and Q-band data obtained by the JVLA in its A-configuration. At an angular resolution
of 0.05 arcsec,  we detected a dense group of hyper-compact (<0.1") radio
sources (HCRs) within the central parsec of the Galaxy. Based on a conservative
15$\sigma$ flux-density threshold, corresponding to 150 $\mu$Jy  at 33 GHz, we
cataloged 64 HCRs with their J2000 equatorial coordinates, flux densities at 33
and 44.6 GHz, angular sizes derived from 2D Gaussian fitting, and spectral index,
$\alpha_{44.6/33}$.  HCR49 is double.

The surface-density distribution, $\Sigma_{\rm HCR}(R),$ shows a local enhancement or a
density bump in the projected radial distance ($R$) range 1.5"--7" superimposed on
a power-law distribution with an index of $\Gamma=1.6\pm0.2$. The steeper profile
of the HCRs relative to that of the nuclear stellar cluster might result from the
concentration of massive stellar remnants by mass segregation.

The 65 HCRs  divide into three spectral sub-types: 38
steep-spectrum ($\alpha_{44.6/33}\le -0.5$),  10  flat-spectrum ($-0.5$ <
$\alpha_{44.6/33} \le$ 0.2), and  17 inverted-spectrum sources ($\alpha_{44.6/33}$ > 0.2).
Our statistical analysis shows that  the distribution of the steep-spectrum
HCRs in Log(S[mJy] can be fitted to a Gaussian with a mean of $\mu=-0.35$
(corresponding to 0.45 mJy) and a standard deviation of
$\sigma=0.22$. We suggest that the steep-spectrum HCRs be
regarded as candidates for a population of stellar remnants acting as
nonthermal compact radio sources powered by accretion onto neutron stars and
stellar-mass black holes, with the accreted matter supplied either by a binary
companion or by a dense portion of the interstellar medium.

The inverted-spectrum HCRs show a rising spectrum towards high frequencies.
Five of the 17 inverted-spectrum HCRs have compact IR counterparts, suggesting
that they are associated with the ionized stellar wind outflows from hot, massive stars.
A portion of the inverted-spectrum HCRs may consist of X-ray binaries in the hard
state, when the self-absorbed synchrotron emission is dominated
by the corona of the compact object. The GC magnetar, SGR J1745-2900,
belongs to the inverted-spectrum sub-type. Based on our analysis
of 91 flux-density measurements of SGR J1745-2900 observed
with the JVLA, ALMA and ATCA, we find that two distinguishable
spectral components contribute to  the averaged spectrum, separated at
the transition frequency $\nu_t\sim 30$ GHz. The cm-component is fitted to
a power-law with a steep spectral index $\alpha_{\rm cm} = -1.5\pm0.6$
while the mm-component shows the inverted spectrum $\alpha_{\rm mm} = 1.1\pm0.2$.
Our consolidation of the spectrum from the interferometer array data
is in good agreement with  earlier results based on single-dish observations \citep{tor2017}.
 
In addition, we reduced 225-GHz ALMA data observed with a FWHM beam
of 0.024"$\times$0.017" on ten individual days within two weeks in October, 2017.
SGR J1745-2900 was detected with a signal-to-noise ratio of $\sim$70 at flux
densities varying widely between 0.79 and 5.60 mJy on day-to-day timescales.
An index of relative variability $\mathscr{RV} \approx $  6 is found
for SGR J1745-2900, which compares with  $\mathscr{RV}\approx $  0.2
for the more moderately variable Sgr A*, derived from the same ALMA observations.

Collecting a total of 161 individual flux-density measurements of
SGR J1745-2900 at radio wavelengths from prior published literature and this
paper, we binned data into 100-day bins along the time axis for both cm- and
mm-components separately,  starting from the outburst on April 23, 2013 (MJD 56406).
The radio light curves with bin-averaged flux density are compared with the X-ray
light curve observed with Chandra \citep{rea2020}.
Except for the appreciable  short-timescale variability at millimeter wavelengths,
the long-term trend at both centimeter and millimeter
wavelengths is a slow, but erratic, decrease in radio power,
in contrast to the smoothly decreasing trend in X-ray luminosity.

Because many HCRs are candidates for being remnants of massive stars, we considered
the origin and fate of such remnants, and speculated on the possible reasons for the
difficulty of finding pulsars there. Neutron stars formed within the relatively highly
magnetized central molecular zone of the Galaxy could themselves inherit sufficient
magnetic flux that they are mostly born as magnetars.  In that case, their ultra-strong
magnetic fields (10$^{14}$ - 10$^{16}$ G), acting on plasma trapped in their magnetospheres, apply
a spin-down torque that causes them to have relatively short lifetimes as detectable pulsars
($10^3~-~10^5$ yr) compared to normal pulsars.

If the inverted spectrum towards high frequencies found in SGR J1745-2900 and 1E1547.0-5408
is a common characteristic of active magnetars, the high-frequency bands (Ka and Q)
of the JVLA and all the ALMA and SMA bands open a practical window for studying magnetars.
Also, dynamical information on the HCRs can be acquired with high-precision proper motion data
that can be obtained with repeated JVLA and ALMA observations.   Such
measurements will supply strong constraints on the candidacy of the HCRs
as massive stellar remnants at the Galactic center.

\acknowledgments
{We are grateful to the anonymous referee and editors for providing 
their valuable 
comments and suggestions. The Jansky Very Large
Array (JVLA) is operated by the National Radio Astronomy Observatory
(NRAO). The NRAO is a facility of the National Science Foundation
operated under cooperative agreement by Associated Universities, Inc.
ALMA is a partnership of ESO (representing its member states), 
NSF (USA) and NINS (Japan), together with NRC (Canada),
NSC and ASIAA (Taiwan), and KASI (Republic of Korea), in cooperation with the 
Republic of Chile. The Joint ALMA Observatory (JAO) is operated
by ESO, AUI/NRAO and NAOJ. This paper makes use of the data from the
following ALMA programs: ALMA\#2018.A.00052.S, ALMA\#2017.1.00503.S and ALMA\#2015.1.01080.S.
This work has been partially supported by NSF grant AST1614782 to UCLA.
The research has made use of NASA's Astrophysics Data System.}

\vskip 40pt

\appendix
\section{Density distribution of HCRs and GCCRs}
The stellar density distribution in the Galactic center 
can be fitted by two power laws \citep[{e.g.,}][]{gen2003, fri2016}, 
with a flat inner component and a steeper component at 
large projected radii \citep{sch2014, sch2018}. Usually, 
the projected surface density can be described by the ``Nuker'' profile with
two slopes \citep{lau1995}:

\begin{equation}
\centering
\Sigma(R) = \Sigma(R_b) 2^{(\beta-\Gamma)/\alpha}\left(R_b\over\/R\right)^\Gamma
\left[1+\left(R\over\/R_b\right)^\alpha\right]^{(\Gamma-\beta)/\alpha)}
\end{equation}
 
\parindent 0pt
where $\Sigma(R_b)$ is the surface density at the break radius $R_b$ that divides 
the profile into two power laws. At a small radius $R\ll R_b$, $\Sigma(R)\sim\/R^{-\Gamma}$, 
usually describing the cusp in the distribution of old stellar population.
For $R\gg R_b$, $\Sigma(R)\sim\/R^{-\beta}$, fitting to the outer power law, which is usually
steeper than that of the cusp \citep{lau1995}. The break radius $R_b$ corresponds 
both to the point at which the slope is the mean of $\beta$ and $\Gamma$ 
and to the radius of maximum curvature of the distribution  in the log$_{10}$($R$)-log$_{10}$($\Sigma(R_b)$) 
coordinate system.

\parindent=3.5mm
 
In this section, we present the algorithm that we developed to derive the distribution of the surface density 
($\rm \Sigma_{\rm crs}$) for the compact radio sources (CRS) as function of the projected 
radial distance from Sgr A*.

\subsection{Algorithm}

First, the radio bright zone (RBZ) is divided into 20 annular zones (AZ), each of 
them bounded by two rings at outer and inner radii, $R(i)$ and $R(i+1)$:

\begin{equation}
\centering
R(i) =2^{-0.5i} R_{\rm RBZ},
\end{equation}
where $i=0, 1, 2 ..., 19~{\rm and}~20$ and $R_{\rm RBZ}$ is 
the outer radius of the radio bright zone (RBZ). 
Fig. A1 shows the 20 annular zones along with the distribution of GCCRs and HCRs 
in the RBZ (Fig. A1a), Sgr A West (Fig. A1b) and the central parsec (Fig. A1c).
The radius of the outer ring, $R(0) = R_{\rm RBZ}$,
corresponds to the largest radius within which we searched the GCCRs with a
10-$\sigma$ cutoff due to the JVLA primary beam at 5.5 GHz. The radius
of the most inner ring, $R(20)=0.55" \approx \theta_{\rm FWHM}$, is approximately the size of FWHM beam at 5.5 GHz.
The zone AZ-10 marks the boundary, within which we found 64 HCRs 
with the JVLA A-array observations at 33 and 44.6 GHz (this paper). 
Outside this boundary, 110 GCCRs are found from the JVLA 
A-array observations at 5.5 GHz \citep{zmg2020}.

Then, we counted the CRS  in each of the AZs by creating a grid function $crs(j,k)$ for 
each of the CRSs, where $k$ is the annular zone ID from 1 to 20 and $j$ is the CRS source ID 
within an AZ. For given a specific grid cell ($j, k$), a unit value is assigned, {\it i.e.} $crs(j,k)=1$. 
Given a  $k$, the AZ-ID, the number of sources in an annular zone is then derived by summing over $j$:

\begin{equation}
\centering
N(k) = \sum_{j} {crs(j,k)}.
\end{equation} 
In addition, we also created  grid functions of 
$s(j,k)$ and $\sigma(j,k)$ for flux-density and uncertainty, respectively.
Corresponding to a specific source ID: ($j, k$), the values of flux density 
and  1-$\sigma$ uncertainty of the CRS
are assigned to $s(j,k)$ and $\sigma(j,k)$. 
The accumulated flux density of CRS in an annular zone can be determined:  

\begin{equation}
\centering
S(k) = \sum_{i=1}^{N(k)}{s(j,k)},
\end{equation}
and

\begin{equation}
\centering
\sigma(k) = \sqrt{\sum_{j=1}^{N(k)} {\sigma(j,k)^2}}.
\end{equation}
We note that, for the GCCRs, $s(j,k)$ and $\sigma(j,k)$ correspond to the mean  and 1-$\sigma$ error of the mean, 
which are determined with the measurements from the three-epoch observations at 5.5 GHz \citep{zmg2020}.
For the HCRs, we used Ka-band flux density and 1-$\sigma$ error at 33 GHz (this paper). 
The area for the $k$-th AZ can be computed:

\begin{equation}
\centering
A(k) = \pi\left[R(k)^2-R(k+1)^2\right].
\end{equation}
Furthermore, we also estimated a flux-density-based CRS counts $N_S(k)$:

\begin{equation}
\centering
N_S(k) = {S(k)\over S_{\mu,5.5~{\rm GHz}}},
\end{equation}
for $k <11$ and 

\begin{equation}
\centering
N_S(k) = {S(k)\over S_{\mu,33~{\rm GHz}}}
\end{equation}
for $k\ge11$. Here $S_{\mu,5.5~{\rm GHz}}$ and $S_{\mu,33~{\rm GHz}}$
are the mean flux-density values for the GCCRs at 5.5 GHz and HCRs at 33 GHz.

Given the sensitivity cutoff and the variability of the CRSs, the direct source counts 
$N(k)$ represents a lower limit of the $k$-th AZ. The flux-density-based CRS counts $N_{S}(k)$  
is usually greater than the direct source counts, {\it i.e.}, $N_{S}(k) > N(k)$.
Thus, using the averaged values of $\displaystyle {\left[N_{S}(k)+N(k)\right]/2}$
we derived the surface density of CRSs as a function of radius from Sgr A*:

\begin{equation}
\centering
\Sigma_{\rm crs}(k) = \displaystyle {\left[N_{S}(k)+N(k)\right]\over 2A(k)}.
\end{equation}
The uncertainty of surface density is estimated as well:

\begin{equation}
\centering
\Delta_{\Sigma_{\rm crs}}(k) = \displaystyle {\left[N_{S}(k)-N(k)\right]\over 2A(k)}.
\end{equation}
where $k=1, 2, 3, $ ... and 20.
Thus, the upper and lower limits in surface density correspond to
the flux-density-based density ($N_{S}(k)/A(k)$) and the actual-source-counted 
density ($N(k)/A(k)$), respectively.

We then calculated the mean radius given the $k$-th annular zone:

\begin{equation}
\centering
R_{\rm AZ}(k) = {{R(k)+R(k+1)}\over 2}, 
\end{equation}
and the uncertainty, or more properly  half of the annular zone width: 

\begin{equation}
\centering
\Delta_{R_{\rm AZ}}(k) = {{R(k) - R(k+1)}\over 2}, 
\end{equation}
where $R(k)$ is defined in Eq.(A1) and $k=1, 2, 3, $ ... and 20.

Fig. A2 shows the distribution of the surface density of CRSs
$\Sigma_{\rm crs}(R)$ as a function of radial distance ($R$) from Sgr A*.
The black symbols represent the data at a large projected radius ($R>13$"), 
determined from the JVLA observations at 5.5 GHz \citep{zmg2020} 
while the symbols for the inner annular zones ($R\le13$") mark the data 
derived from the Ka-band observations at 33 GHz. We note that the  CRS 
counts, as well as the surface density $\Sigma_{\rm crs}$ (light-grey symbols in Fig. A2),  
may be underestimated within the inner annular zone (AZ=11 - 20) due to  
the fact that the data HCRs observed at 33 GHz corresponds to a higher effective 
cutoff limit  at 5.5 GHz since most of the HCRs have a steep spectrum. 
With an averaged spectral index $\alpha_{\rm HCRs} = -0.42\pm0.12$
using all 64 HCRs' data, we converted the 33-GHz flux densities to   5.5-GHz. The equivalent 
cutoff limit for the inner AZs is  about twice as high as  the limit 
searched for GCCRs in the RBZ \citep{zmg2020}. 
So, we multiplied the surface density of the inner AZ(k$\ge$11) by a correction factor: 

\begin{equation}
\centering
\eta = (5.5/33)^{\alpha_{\rm HCR}} \left(\sigma_{33}\over\sigma_{5.5}\right)\approx2.4,
\end{equation} 
assuming that the HCR counts
are inversely proportional to the cutoff value in flux density. 
The corrected surface density is then computed: 

\begin{equation}
\centering
\Sigma^*_{\rm crs}=\eta \Sigma_{\rm crs}
\end{equation}
as indicated with the
red symbols  which appear to be  aligned  better with the  
surface density data of the outer radii obtained from the 
observations at 5.5 GHz.    

\renewcommand{\thefigure}{A1}
\begin{figure}[htp]
\centering
\includegraphics[angle=0,width=99.5mm]{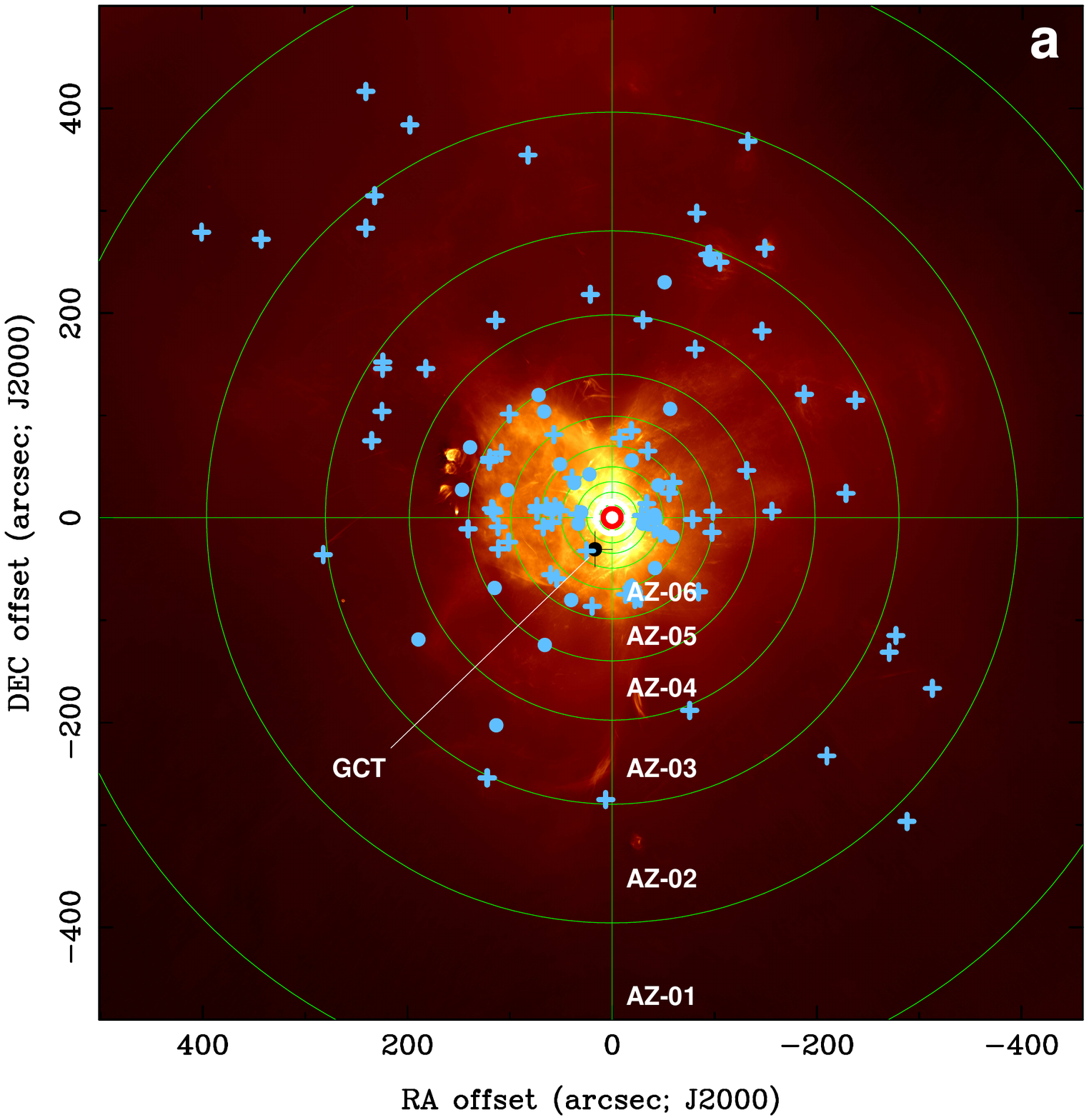}
\includegraphics[angle=0,width=99.5mm]{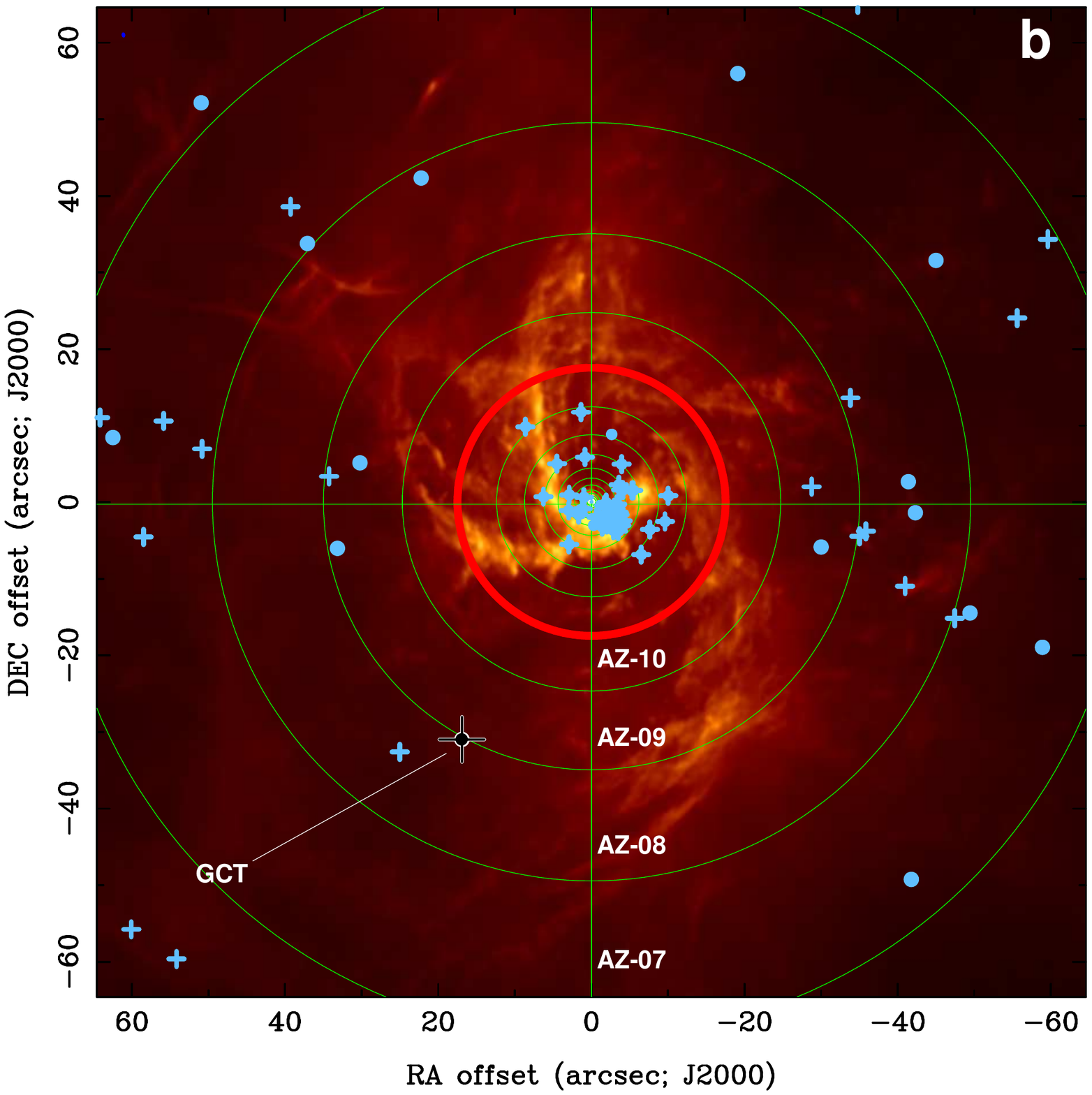} 
\includegraphics[angle=0,width=99.5mm]{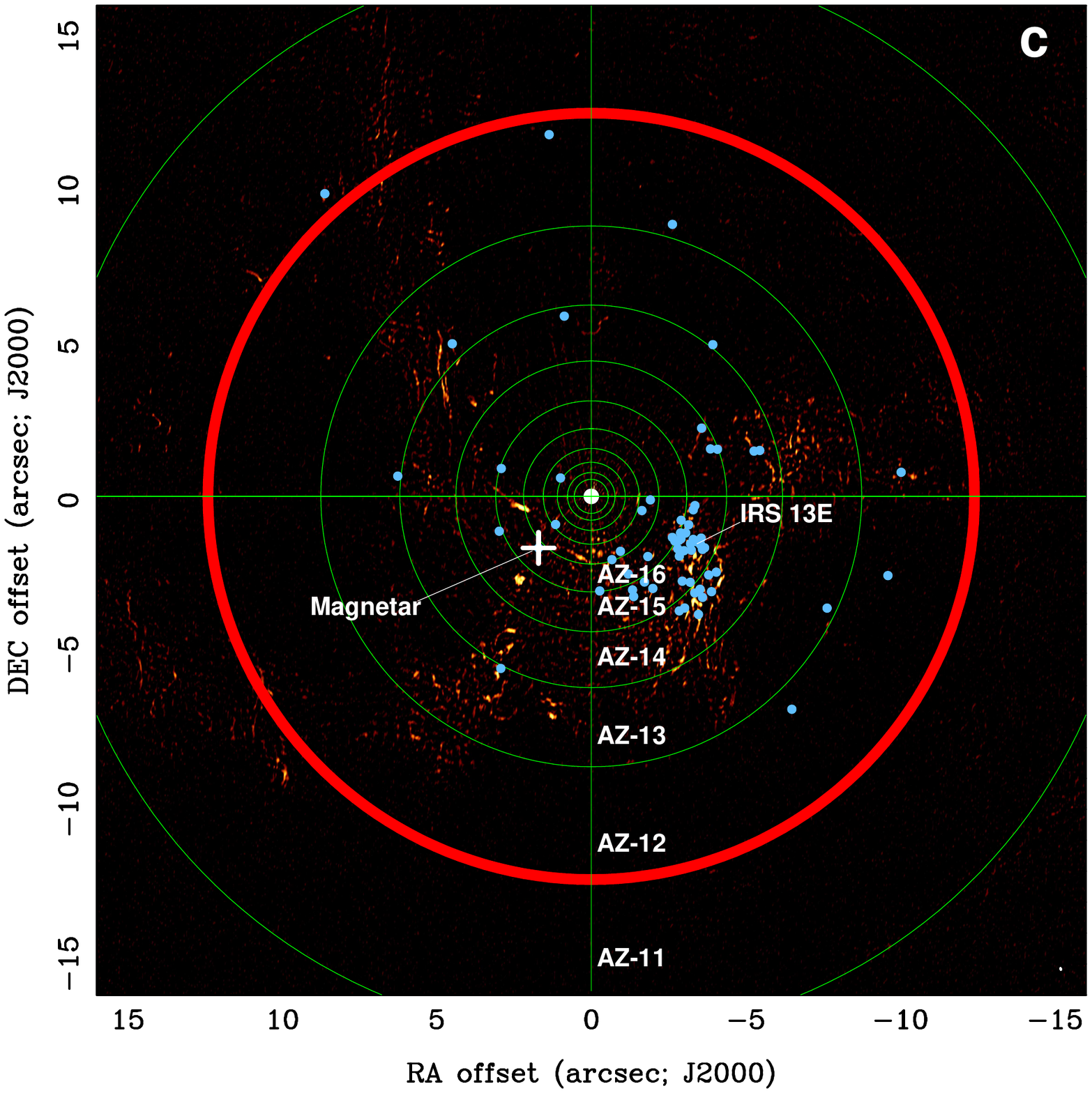}
\caption{The radio bright zone (RBZ) is divided into 20 annular zones (AZ) 
along the radial direction from Sgr A*.
Spatial distribution of GC compact radio sources or GCCRs 
(light blue symbols, crosses for variables and dots for non-variables) in annular zones is displayed over 
the RBZ (AZ 1 - 6) (top panel: Fig. A1a) and
SgrA West (AZ 7-11) (middle panel: Fig. A1b). The HCRs (light blue dots)
are distributed in the central parsec (AZ 12-20) (bottom panel: Fig. A1c). 
The thick red ring marks the 11-th ring (Eq. A2),
$R(11)=13$", $\sim0.5$ pc. In Figs. A1a and A1b, the symbol of a black cross-dot marks the Galactic center 
transient, the GCT \citep{zha1992}. The magnetar J1745-2900 and IRS 13E are labelled in Fig. 1Ac.
}
\end{figure}

\subsection{Power-law fit to the $\Sigma_{\rm crs}$ distribution}

The surface-density $\Sigma_{\rm crs}$ data at large radii (from AZ01 to AZ09) can be described 
by a power law while the  data at small radii (from AZ10 to AZ18) show  
a bump ($\Sigma_{\rm crs}$-bump), 
indicating a locally enhanced source density $\Sigma_{\rm crs}$ 
at the  radial distance range of  
$R=1.5$"$- 7.0$" (from AZ-17 to AZ-14) 
superimposed on a slope.  
We made a linear regression fit to a function: 

\begin{equation}
\centering
Log_{10}(\Sigma_{\rm crs}) = a_{\rm GCCR} -\beta\/Log_{10}(R),
\end{equation}
using  the outer nine AZ's data points (GCCRs) with a weight of 
$wt(i) = \Delta^{-2}_{\Sigma_{\rm crs}}(k)$.  
The best fit  derived from the least squares regression
(LSR), as indicated by a straight blue line in Fig. A2, 
gives a  power law function $\Sigma_{\rm crs}\propto\/R^{-\beta}$ ($R>13$") 
with $\beta=2.0\pm0.1$ and $a_{\rm GCCR}=4.3\pm0.2$.

For the inner eight annular-ring data (HCRs), we excluded the four data points 
that are involved in the  $\Sigma_{\rm crs}$-bump (from AZ14 to AZ17) and fit the 
remaining four data points with a linear function:

\begin{equation}
\centering 
Log_{10}(\Sigma_{\rm crs}) = a_{\rm HCR} -\Gamma\/ Log_{10}(R).
\end{equation}
Based on a least-squares fitting, we derived $\Sigma_{\rm crs}\propto\/R^{-\Gamma}$ ($R\le13$")
with $\Gamma=1.6\pm0.2$ and $a_{\rm HCR}=3.7\pm0.2$, 
as indicated with the 
straight orange line in the logarithmic plot (Fig. A2a). The $\Sigma_{\rm crs}$-bump
is characterized by
the high surface-density zones AZ15 and AZ16; the former  contains the GC magnetar
J1745-2900 and the latter is associated with IRS 13E. We note that
the surface brightness in the diffuse Br$\gamma$ narrow band filter image
(Fig. 6a of \cite{sch2018})
also shows a bump in the range of radius between $R=1.5$" and $7.0$", suggesting that 
the counts of CRSs in the radial range  may be partially caused by the local
hyper compact HII sources associated with younger massive stellar objects.
However, excluding the flat-spectrum HCRs, the $\Sigma_{\rm crs}-$bump is also
present (see Fig. A2b). We further checked the distribution only with  the steep-spectrum
HCRs and found that the $\Sigma_{\rm crs}-$bump is still present in the radial range between $R=1.5$" and $7.0$"
(see Fig. A2c).

Due to a small number of data points and at least four free parameters in the
``Nuker'' model of Eq(A1), we made no attempt to fit the entire curve 
for deriving the break radius  $R_b$.
However, from the intersection between the two straight lines of Eq(A15) and Eq(A16), 
we found an approximate break radius $R_b\approx 30$", or 1.1 pc, which appears to correspond to 
the radius of the ionized region. 
Therefore, in the main text, we  use $R_b=30$" as the break radius 
that separates the two distinctive power-law distributions. 

\renewcommand{\thefigure}{A2}
\begin{figure}[htp]
\includegraphics[angle=0,width=90.5mm]{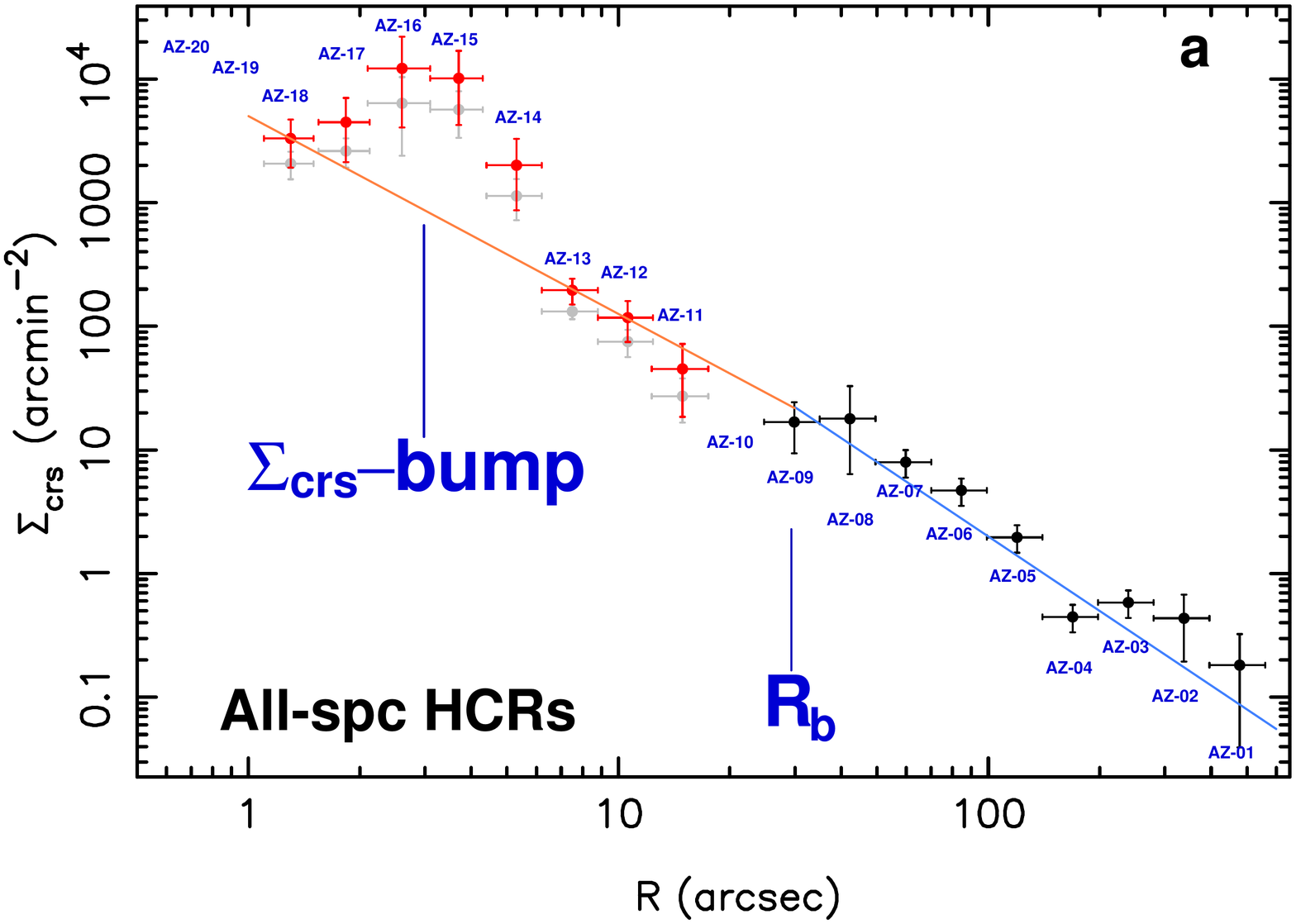} \\
\includegraphics[angle=0,width=90.5mm]{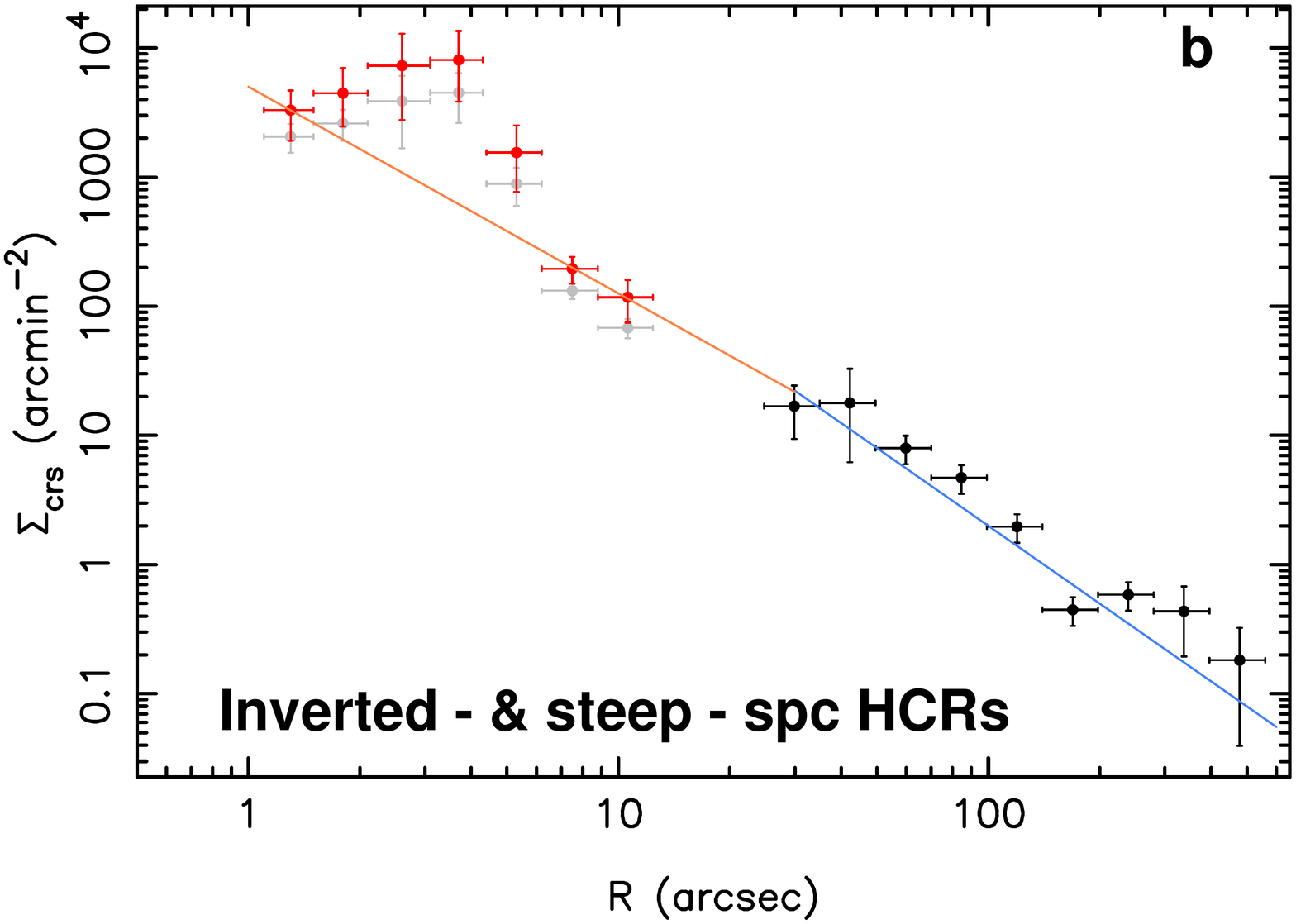}
\includegraphics[angle=0,width=90.5mm]{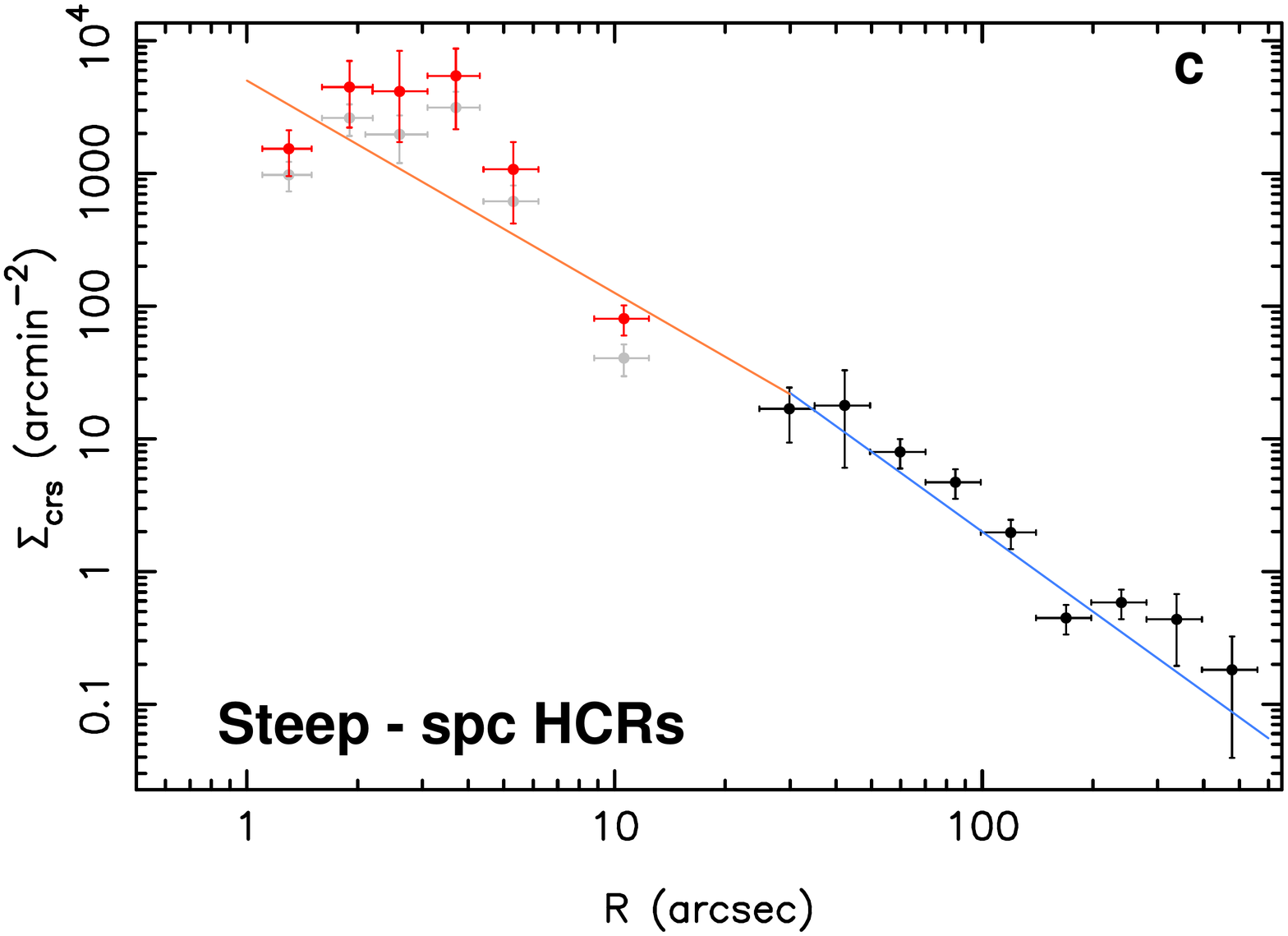}
\caption{Radial distribution of surface density of radio 
compact sources detected at the Galactic center. The red points are the surface density data 
$\Sigma^*_{\rm crs}$ 
corrected for a higher equivalent sensitivity cutoff at 5.5 GHz by multiplying a 
correction factor $\eta =2.4$ (see Eqs. A13 and A14) to the original 33-GHz surface 
density data $\Sigma_{\rm crs}$ (light-grey points) derived from the 33-GHz 
observations of the HCRs (this paper). The black points mark the data of GCCRs
observed at 5.5 GHz. In the annular zones around $R=3"$, a bump in 
$\Sigma_{\rm crs}$ is present. The orange line shows the best fitting to 
the power-law function with  $\Gamma=1.6\pm0.2$ for the inner region; and
the blue line indicates the least squares fitting to the steeper power law with  
$\beta=2.0\pm0.2$ for the outer region. The break radius $R_b$ is 
$\sim30"$, or $\sim1.2$ pc. Three panel show the surface-density distributions  
for the three combinations of spectrum classes of HCRs: 
(a) inclduing all the HCRs, 
(b) excluding the flat-spectrum HCRs, and 
(c) only including inverted HCRs.
}
\end{figure}

\end{document}